\documentclass{IEEEcsmag}

\usepackage[colorlinks,urlcolor=blue,linkcolor=blue,citecolor=blue]{hyperref}

\usepackage{upmath}
\usepackage[normalem]{ulem}

\jvol{XX}
\jnum{XX}
\paper{8}
\jmonth{May/June}
\jname{Computer Graphics and Applications}
\pubyear{2023}

\usepackage{xcolor}
\usepackage{multirow}

\newcommand{\gr}[1]{\textcolor{black}{#1}}

\begin{document}


\title{MAGES 4.0: Accelerating the world's transition to VR training and democratizing the authoring of the medical metaverse}

\author{Paul Zikas}
\affil{ORamaVR SA, University of Geneva}

\author{Antonis Protopsaltis}
\affil{ORamaVR SA, University of Western Macedonia}

\author{Nick Lydatakis, Mike Kentros, Stratos Geronikolakis, Steve Kateros, Manos Kamarianakis, Giannis Evangelou, Achilleas Filippidis, Eleni Grigoriou, Dimitris Angelis, Michail Tamiolakis, Michael Dodis, George Kokiadis, John Petropoulos}
\affil{ORamaVR SA, University of Crete, ICS-FORTH}

\author{Maria Pateraki}
\affil{ORamaVR SA, National Technical University of Athens, ICS-FORTH}

\author{George Papagiannakis}
\affil{ORamaVR SA, University of Crete, ICS-FORTH}


\begin{abstract}
In this work, we propose MAGES 4.0, a novel Software Development 
Kit (SDK) to accelerate the creation of collaborative medical 
training applications in VR/AR. 
Our solution is essentially a low-code metaverse authoring platform for 
developers to rapidly prototype high-fidelity and high-complexity medical simulations.
MAGES breaks the authoring boundaries across extended reality, since networked participants
can also collaborate using different virtual/augmented reality as well as mobile and desktop devices, in the same metaverse world. 
With MAGES we propose an upgrade to the outdated 150-year-old master-apprentice medical training model. Our platform incorporates, in a nutsell, the following novelties: a) 5G edge-cloud remote rendering and physics dissection layer, 
b) realistic real-time simulation of organic tissues as soft-bodies under 10ms, 
c) a highly realistic cutting and tearing algorithm, 
d) neural network assessment for user profiling and, 
e) a VR recorder to record and replay or debrief the training simulation from any perspective.
\end{abstract}

\maketitle

\chapterinitial{The medical metaverse}, despite the inflated expectations, is steadily, albeit quietly, being created \cite{METAVERSE}.
Along with it, many technical questions remain, including ``who will build the medical metaverse and how?''
Building such an ecosystem from few stakeholders would require 
significant effort involving tasks of tremendous complexity, unless the metaverse authoring process is decantralized and the tools to create it are democratized in the hands of the actual content creators.

The extended pandemic crisis highlighted the need for effective medical
training, along with the inadequacy of the 
150-year-old surgical training model \cite{150YearOldModel}.
Computational medical science 
aims to 
accelerate world’s transition to VR medical training in 
the metaverse \gr{\cite{METAVERSE}} and empower medical professionals to 
enhance their proficiency and ultimately improve patient 
outcomes \cite{HooperTrial, EPICSAVE, VITAWIN, UsingXR}. 

To serve the causes above, we present 
MAGES 4.0, the medical VR industry's first 
Software Development Kit that allows rapid 
prototyping of any shared, 
collaborative networked medical training in VR, in a fraction of time and cost.
It draws its robustness from the 20 years of academic Research \& Development,
incorporating the latest advancements into a novel medical VR/AR 
software development kit.

\begin{figure*}
\centerline{\includegraphics[width=35pc]{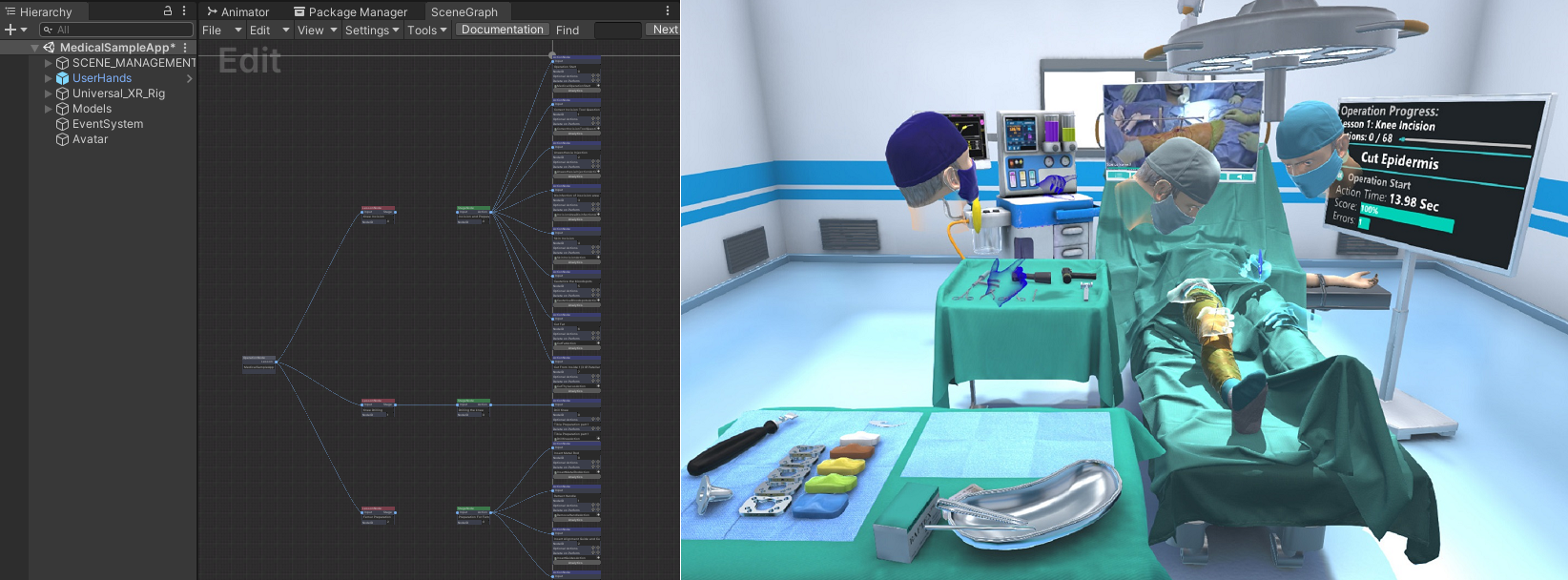}}
\caption{(Left) The visual scripting editor. (Right) A surgeon with two students performing a collaborative Total Knee Arthroplasty as generated using MAGES, in Unity3D.}
\label{fig:scripting}
\end{figure*}

\begin{figure}
\centerline{\includegraphics[width=18.5pc]{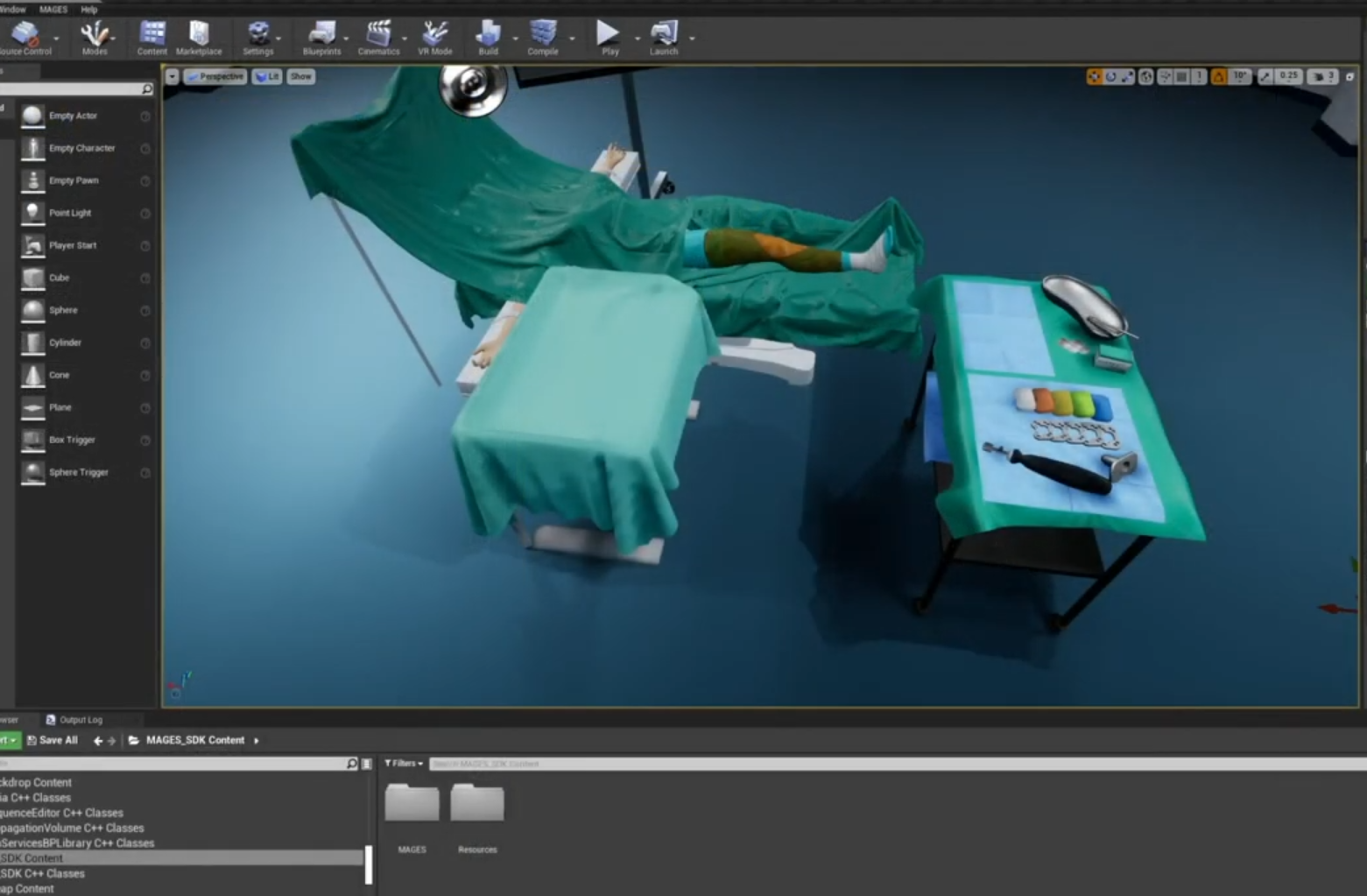}}
\caption{Developing a Total Knee Arthroplasty with MAGES in Unreal engine.}
\label{fig:TKAinUnrealEngine}
\end{figure}

The computational results achieved with MAGES SDK exceed those typically reached by large teams of domain expert developers of similar engines. 
Being layered on top of existing game engines, such as 
Unity3D and Unreal Engine (see Fig.~\ref{fig:scripting}-Left and \ref{fig:TKAinUnrealEngine}), 
it brings a low-code virtual world authoring platform  
to developers with even a moderate knowledge of these engines. 
\gr{This allows a small team, of a one developer and one designer, as in our latest use case (see Case Studies section), to create a complete VR medical training simulation even within two weeks.}

\section{PREVIOUS WORK}
In the past years, a variety of metaverse authoring tools have been 
introduced \cite{coelho2022authoring}, tailored to particular research problems and application verticals, most of 
them presented as proof of concept. 
A great number of platforms have been designed for entertainment/multimedia systems \cite{nebeling2020xrdirector}, education/vocational training \cite{cassola2021novel, VRADA1}, Cultural heritage, and Medical Training \cite{MAGES3, CVRSB, EPICSAVE, VITAWIN, UsingXR}. Each simulation has various strengths, but because they were often designed with specific use cases in mind, each is also limited in different ways. 
Most such VR systems provide high-fidelity photorealistic scenes \cite{nebeling2020xrdirector}, 
user immersion and presence that allow human-centric interaction 
in VR \cite{zhang2020flowmatic}. Existing simulation environments are 
differentiated as they combine VR training functions 
in various domains, while others, only very few of them \cite{MAGES3}, 
provide realistic physics simulations on surfaces with rigid body, 
soft body, cloth, and fluids. Innovative multi-user collaborative virtual environments (VE) \cite{UsingXR, EPICSAVE, nebeling2020xrdirector, MAGES3} provide a great asset in all fields, 
a fact that was especially highlighted during the COVID-19 pandemic. 

Few of the above VE authoring tools \cite{cassola2021novel, MAGES3} 
allow reuse/import of content/assets that impacts the tool’s 
usability and effectiveness, as it enables designers to 
significantly reduce their effort, and generate custom 
user-experience that otherwise would be impossible. 
In that respect, some novel, innovative, and interoperable 
authoring platforms allow users with non or limited programming knowledge to design a VE fast and easy \cite{MAGES3}, 
setting early the basic and essential requirements 
for the building of the metaverse \cite{coelho2022authoring}. 
In that respect, VR software design patterns \cite{SCRIPT} is a 
momentum-gaining approach, with already concrete results,
that will be utilized in the building process of the 
next-generation VR training applications for the metaverse. 

\section{MAGES 4.0 INNOVATIONS}
The \gr{innovations, presented in this work, 
extend significantly our previous work \cite{MAGES3}, and thus transform 
our SDK to a powerful low-code VR authoring platform, that allows 
developers to produce high-fidelity and high-complexity VR simulations fast and effectively, 
empowering the medical metaverse creation.} MAGES SDK is 
named after its most unique features:

\textbf{Multiplayer With GA Interpolation (M):} 
A custom Geometric Algebra (GA) interpolation engine 
allows up to 300 simultaneous users in the same session 
of the VE. To the best of our knowledge, this is the 
highest number of active concurrent users in the same scene, 
whereas state-of-the-art applications offer only a few tens 
of concurrent users with only one active and rest being 
spectators. This remarkable achievement is accomplished by 
the transmission over the network layer of GA-based 
representation forms of the VR scene transformation data, 
which support higher data compression rates. 
The use of such alternative forms by MAGES reduce the 
required data that must be sent over the network by up 
to 33\%, whereas the reduced amount of data sent in these 
forms yield visually better interpolated results, with 
the overall Quality of Experience
being sufficient even when the network quality, in 
terms of bandwidth, seriously deteriorates \cite{LESSisMORE}. 

\textbf{Analytics (A):} 
A powerful advanced analytic engine system allows 
tracking and visualization of the student's progress. 
As this feature enables scoring 
per user action in a simple manner, it provides an 
easy way for the tutor or the 
students to quickly inspect and identify the parts of the training performance. A main insight is that as main recorded data are based on GA, their storage, retrieval or interpolation is much more efficient than existing methods \cite{DL3}.

\textbf{Geometric Algebra Deformable Animation, Cutting and Tearing (G):} 
The under-the-hood GA engine used to solve the animation equation is 
responsible not only for model deformation but for a series of 
features that exploit it. Specifically, the ability to perform 
cuts, progressive tears and drills on skinned soft-body meshes is now feasible 
in real-time, with increased realism suitable even for demanding VR/AR immersive applications (under 10ms frame rendering time) \gr{\cite{ProgressiveTearing}}. 

\textbf{Editor with Action Prototypes (E):} 
An incorporated visual scripting editor 
along with our custom VR simulation design patters
as basic building blocks, called \emph{Actions}, are used to rapidly 
accelerate content creation for VR/AR simulations. 
Based on software design patterns \cite{SCRIPT}, they enable creation of medical training operations, at a fraction of the time and cost, against current practices and standards.

\textbf{Semantically Annotated Deformable, Soft and Rigid Bodies (S):}
Towards a highly realistic recreation of a virtual surgery operation, 
we have designed and developed a novel particle system, suitable for 
real-time elasticity simulations of human tissues and organs in VR/AR \gr{\cite{ProgressiveTearing}}.

\gr{
The above novel features of MAGES 4.0 were significantly enhanced, compared to our previous work 
\cite{MAGES3}, in order to provide higher fidelity for incisions, 
increased performance for Soft bodies, 
extended capabilities for ML Analytics (connection with 
the built in VR-Recorder),
ability to handle more concurrent users via the GA networking 
layer and support for a larger set of actions for the Scripting 
Editor; thus paving the way for MAGES 4.0 to become a complete low-code metaverse authoring tool.
}

\section{HOW MAGES WORKS IN 5 STEPS}

In this section we present the five main steps, a developer should follow, to create a VR training simulation:

\begin{enumerate}
  \item \textbf{Design training storyboard:} Utilizing our visual scripting editor, developers create the steps of each scenario.
  \item \textbf{Create virtual assets:} Gather all the medical 3D content (tools, human anatomy, etc.).
  \item \textbf{Author training Actions:} Generate the Action scripts using our Action prototype design patterns. Developers create programmable Actions along with analytics and network behavior using the embedded authoring tools.
  \item \textbf{Build the medical VR training simulation:} Build and deploy the executable application to a wide range of supported headsets/platforms, operating systems and portable/desktop devices.
  \item \textbf{License the simulation:} Optionally, connect the 
   the training simulation to an analytics server, using 
 a MAGES SDK cloud management license.
\end{enumerate}

Code reusability and prototyping are two major principles in 
software architecture. The structure of software systems and 
the communication between its modules is described through 
software design patterns. Software design patterns are reusable solutions for common programming problems that often occur 
during software development \cite{PATTERNS}. \gr{In \cite{CGI2}, 
the creation of prototyped behaviours in VR with a system to 
manage and visualize interactive actions was simplified. 
The system is able to analyse user behaviors in 
VR to export highly accurate maps of user engagement that can
be used later for in-VR action understanding.}

\section{ACTION PROTOTYPES}

Especially in training simulations, developers need to 
implement highly interactive behaviors for the students to follow. 
For this reason, we introduce Action prototypes, as novel software 
design patterns, for low-code behavioral tasks in training scenarios. 
\gr{The novelty in such an approach lies in the fact that these software patterns are tailored specifically for authoring of VR/AR training experiences.} \cite{SCRIPT} e.g. "Insert action", "Question/Answer action", "Tool Action" etc. 
We classified the majority of the physical tasks into programmable 
and easily extendable code patterns. 
We implemented basic behaviors, like the insertion, removal, 
usage, or even cognitive behaviors, like questions into 
separate programmable entities. Developers can inherit those 
entities and develop their own Actions with a few lines of code.

To minimize the need of coding, we developed an authoring tool 
to automatically generate networked-ready, analytics-ready Action scripts. 
Developers can create such Action scripts by selecting the 
necessary physical objects for each task (e.g tools, 
specific parts of the body) and subsequently the system 
will automatically generate an Action script. 
This script contains the basic-default behavior of the 
particular object, easily extendable by mounting extension-scripts.

\section{THE TRAINING SCENEGRAPH}
A training scenario contains a number of carefully defined 
steps, in a sequential or multi-path manner. 
MAGES \gr{usage depends on an underlying} training \emph{Scenegraph}, 
a highly dynamic, acyclic graph representing the training 
scenarios. Each node defines an \emph{Action}, a specific task 
to be completed by the trainee. 

\gr{A training} scenegraph is not just a static tree, it’s a 
dynamic graph \cite{SCRIPT}. An educational scenario can lead to multiple 
paths according to user’s actions and decisions. 
To accommodate this need, the training Scenegraph can generate 
new paths and deviate from the originally intended path while 
the user explores the training scenario.

To rapidly accelerate the content creation we built a visual 
scripting authoring tool. This is a low-code system that 
enables easy authoring of Actions while presenting the scenario 
from a higher level perspective. The visual scripting editor 
\gr{is able to visualize the training scenegraph and, as it 
consists of interactive nodes, allows the user-developer to 
modify it. This feature} is able to generate training simulations 
in a future proof \gr{implementation-agnostic} way, \gr{via a user friendly GUI} (see Fig.~\ref{fig:scripting} (Left)).

\gr{Although VR medical training improves patient outcomes 
\cite{HooperTrial}, VR content creation is a lengthy and 
costly process\footnote{https://roundtablelearning.com/cost-of-virtual-reality-training-full-vr-2020}, 
as the development of a small 
simulation requires four man-months for one designer and 
one developer. The use of the visual scripting editor allows the 
authoring of a small VR medical simulation by the same personnel in 
just 14 days (see \nameref{sidebar:case_studies} for such examples), 
therefore it is 8 times faster and cost-efficient than SoA methods.}

\section{ANALYTICS EDITOR}
User assessment is crucial in medical simulations as students 
can identify their skills, while teachers can note any difficulties 
or points of improvement. 
For this reason we integrated into MAGES an analytics system based on \cite{DL3}
to assess and track the progress of each individual. 
In our training scenarios, the corresponding medical subject-matter-experts are the ones that  
direct which specific performance parameters to track throughout the medical operation. 
Those can vary from "wrong angle during incisions" or "tool placements", 
to "time constrained" decisions or even "contamination" issues. 

MAGES features a novel analytics system to configure and 
present the assessment data. We built an easy to use interface 
on top of our visual scripting engine that enables analytic functionality to our Actions, via minor modifications. 
For this reason, we introduced the scoring factor, a 
component that tracks data from objects or from the student's 
actions. Some of the scoring factors we implemented are 
velocity scoring factors (to track if a user is moving a 
fragile object too fast), error collider scoring factors 
(to check for possible contamination), as well as angle 
scoring factors (to check the proper placement of implants). 
Custom scoring factors are also supported allowing developers 
to extend the existing classes and create their own factors. 
Developers can assign multiple scoring factors in each Action. 
The final score for each Action is calculated as a weighted 
average of all assigned scoring factors.

After each simulation, users can visualize their analytics 
from within the VR/AR environment. Analytic reports are also 
uploaded automatically to our online portal allowing 
supervisors and teachers to view their students' progress. 
The implementation of the portal upload feature is as simple as  
a single API call in the MAGES configuration file.

\section{AUTOMATIC HAND POSTURES}
Interactive virtual characters are nowadays common 
place in VR/AR applications. Designing a virtual human-like 
hand which is able to touch and grasp objects with realistic 
hand and finger adjustments, often trying to imitate the human 
brain way of reasoning, is a matter of necessity. 
Grasping hand animation itself includes several 
problematic aspects that make finding a satisfying solution 
extremely hard. Especially concerning human-like hands, 
the main challenge is trying to imitate the human-brain 
way of thinking and replicate the unconscious movements 
a person makes to grab an object. \gr{Although it’s true that 
a 3D physics engine can quite easily manage collisions 
and anatomical constraints, the real challenge is to 
design and implement a controller that supervises all the 
motors of the hand \cite{Interaction2}. 
Predefined animation can’t be used in general on a 
physical body, therefore even producing a simple movement 
requires particular care.}

Reducing the number of possible grasp movements is 
fundamental. The applied methodology should also respect 
the physiology of the human hand. 
This applies to anatomical joint limits, angle limits due 
to tendon links between fingers (dynamic constraints) 
as well as angle limits that force a natural posture.

\begin{figure}
\centerline{\includegraphics[width=18.5pc]{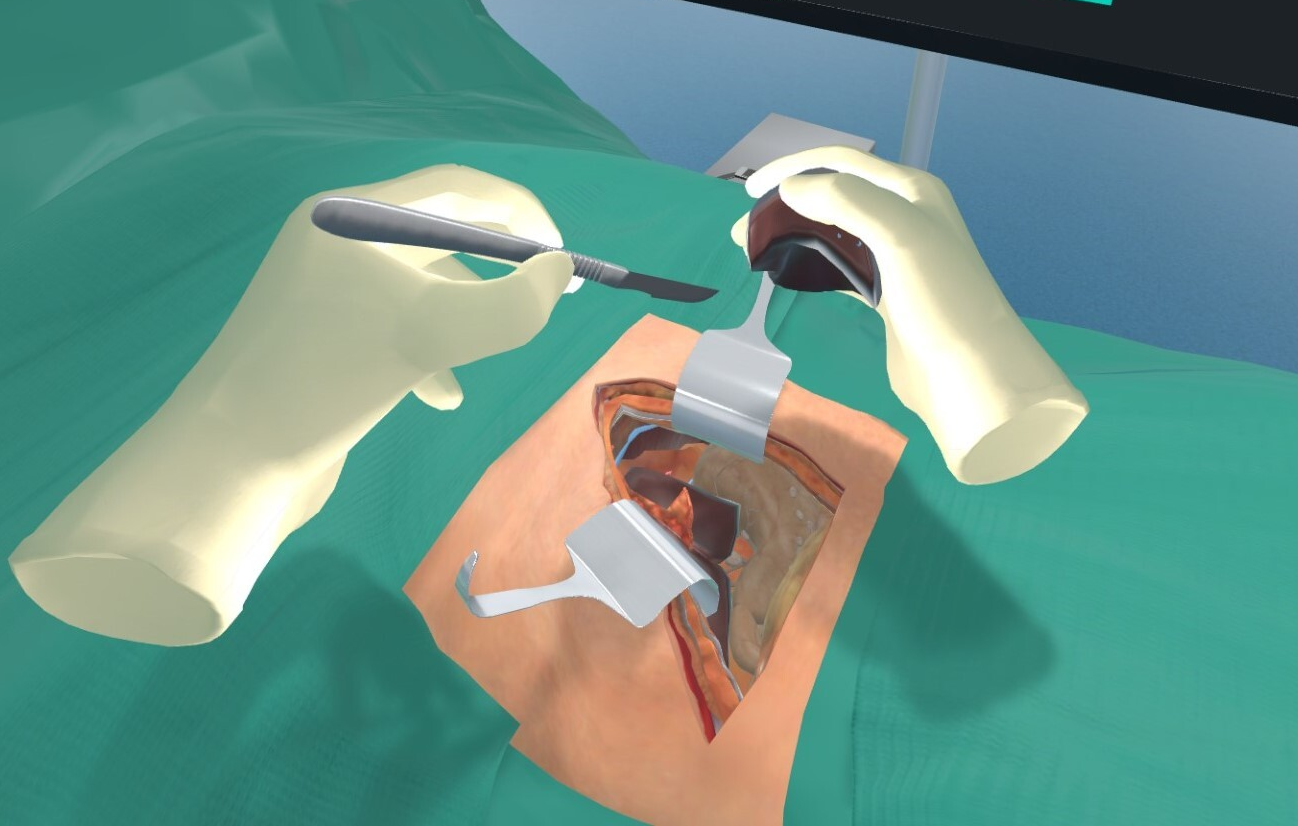}}
\caption{A student has just dissected the left part of the liver, 
during a hemihepatectomy operation. MAGES provides a realistic simulation 
of hand postures when grasping various objects.}
\label{fig:automaticHandPostures}
\end{figure}

In MAGES we designed an algorithm for intuitive object 
grasping with easy configuration and high accuracy, 
mimicking the human way of reasoning. 
This algorithm is flexible enough to support hand 
structures having an arbitrary number of fingers and 
an arbitrary number of phalanges for each finger. 
Given the hand bones structure, the position of the character 
and an object, the algorithm finds the most suitable 
grasping position for the character’s hand on that object. 
Grasp poses are generated by a combination of a generic 
grasp movement and a target object that the fingers collide with. 
Movements are stored using only an initial pose and a final pose, 
so that the algorithm can generate an arbitrary number 
of interpolations between them. The algorithm runs in 
real-time without imposing a performance overhead to the 
main program, even with complex objects 
(see Fig.~\ref{fig:automaticHandPostures}).

To simulate grasp poses, we implemented a procedure that, 
based on a starting hand position and a grasp movement, generates 
the grasping pose, by considering the object and 
computing the final positions and rotations for all bones. 
When the user grabs an object, starting from the root joint of 
the hand, we interpolate the joints to reach the final pose. 
If a bone gets in contact with the object, it is excluded 
from the following calculations. This algorithm runs recursively 
for each joint, resulting in a firm grip. 

\section{EXTENDED REALITY, MULTI-MODAL DEVICE SUPPORT}
\begin{figure}
\centerline{\includegraphics[width=18.5pc]{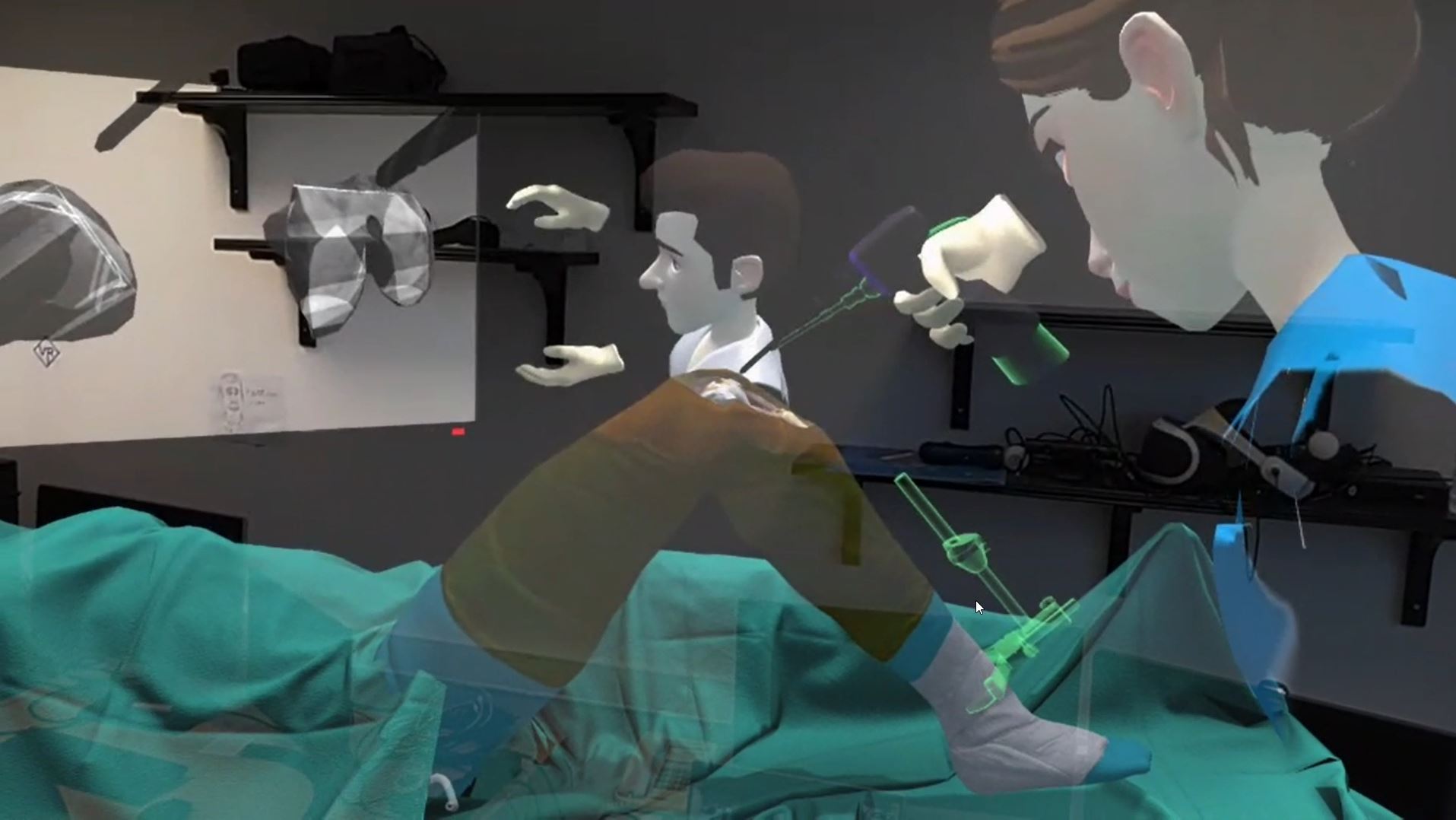}}
\caption{Collaborative Total Knee Arthroplasty simulation between a Magic Leap (point of view) and an oculus quest 2 headset (the medical assistant).}
\label{fig:collaborativeARandVR}
\end{figure}

Access to virtual worlds should be permitted in a seamless way, 
regardless the nature of the XR device used. 
For this reason, MAGES is designed to allow virtual sessions
for both VR and AR headsets, mobile and desktop devices, without requiring complex 
configurations and additional work from the developer,  
allowing such users to collaborate in the same session 
(see Fig.~\ref{fig:collaborativeARandVR}). 
In this regard, if a developer needs to build an application 
for an AR headset (e.g HoloLens, Magic Leap etc), the only 
additional work required is to import the respective external SDK. 
Our universal XR camera inter-operates with any HMD 
regardless of the targeted reality\gr{, by automatically 
culling unnecessary VE object (e.g., operating room walls) 
when rendering for AR device, breaking the boundaries 
between different XR technologies, that up until recently could
not be combined}.


\section{EDGE-CLOUD REMOTE VISUAL RENDERING}
Although the VR hardware landscape evolves rapidly, 
with SoC solutions that aim to reduce the performance gap 
to high-end desktop GPUs, still standalone/untethered VR solutions 
are less capable in supporting high quality strong 
interactive VR services, due to their reduced GPU 
capabilities and battery life. 
This favors the exploitation of software solutions that offload 
computationally intensive tasks from end-devices to cloud/edge resources to support processing and storage. 
\gr{Recently, the NVIDIA's CloudXR solution has emerged for 
remote rendering and streaming from OpenVR applications on 
a remote server-cloud or edge, attained via a client 
application dependent on the HMD and the OS installed on the 
client's local machine. In our approach, we overcome the dependencies 
that regard the device's proprietary API's, thus 
supporting cross-platform access to VR devices. This is accomplished 
by handling rendering and streaming inside the offloaded 
application based on Unity engine. }

A signaling server is further integrated to complete the 
handshake and the \gr{Interactive Connectivity Establishment\footnote{For more information, also refer to the RFC 8445 standard.}} 
candidate \gr{exchange} 
between the HMD and the offloaded application. 
The signaling server 
establishes a communication channel \gr{by a TURN server hosted inside the offloaded application},
and decides which encoding format is the most 
appropriate to use for the video stream. 

\gr{Although the Unity XR SDK is exploited for rendering, several 
custom modifications had to be applied.} 
One major road-blocker for VR offloading using Unity3D Render 
Streaming is that Unity3D does not support stereo render-to-texture. To cope with this limitation,  we re-engineered the render 
streaming by using two complete Unity3D cameras in the 
scene. Each camera included customized settings for the left 
and right eye in order to render separate textures for each eye, 
which are then combined. Furthermore, a 
custom data stream is used to send transformation and 
controller data from the HMD to the offloaded application. 
With the simulation of two Unity3D cameras the frame-rate between the offloaded component and the HMD 
\gr{reaches 60 fps} on average. 
The offloaded application runs inside of a Windows virtual machine 
(VM) on a linux host. By using KVM with GPU-passthrough, performance 
can come close to bare-metal speeds. Additionally, a custom TURN 
server is hosted inside the VM, which allows WebRTC to 
consistently establish communication with the client, even if they 
happen to be behind a NAT.

\section{DISSECTED EDGE-CLOUD REMOTE PHYSICS ENGINE}
A typical monolithic game engine pipeline involves the execution 
of physics related calculations, performed on CPU, alongside with 
scene rendering calculations, performed on CPU/GPU. Both these 
heavy computations are either performed on high-end untethered HMDs, 
or on VR-ready PCs with a tethered HMD.

\begin{figure}
\centerline{\includegraphics[width=18.5pc]{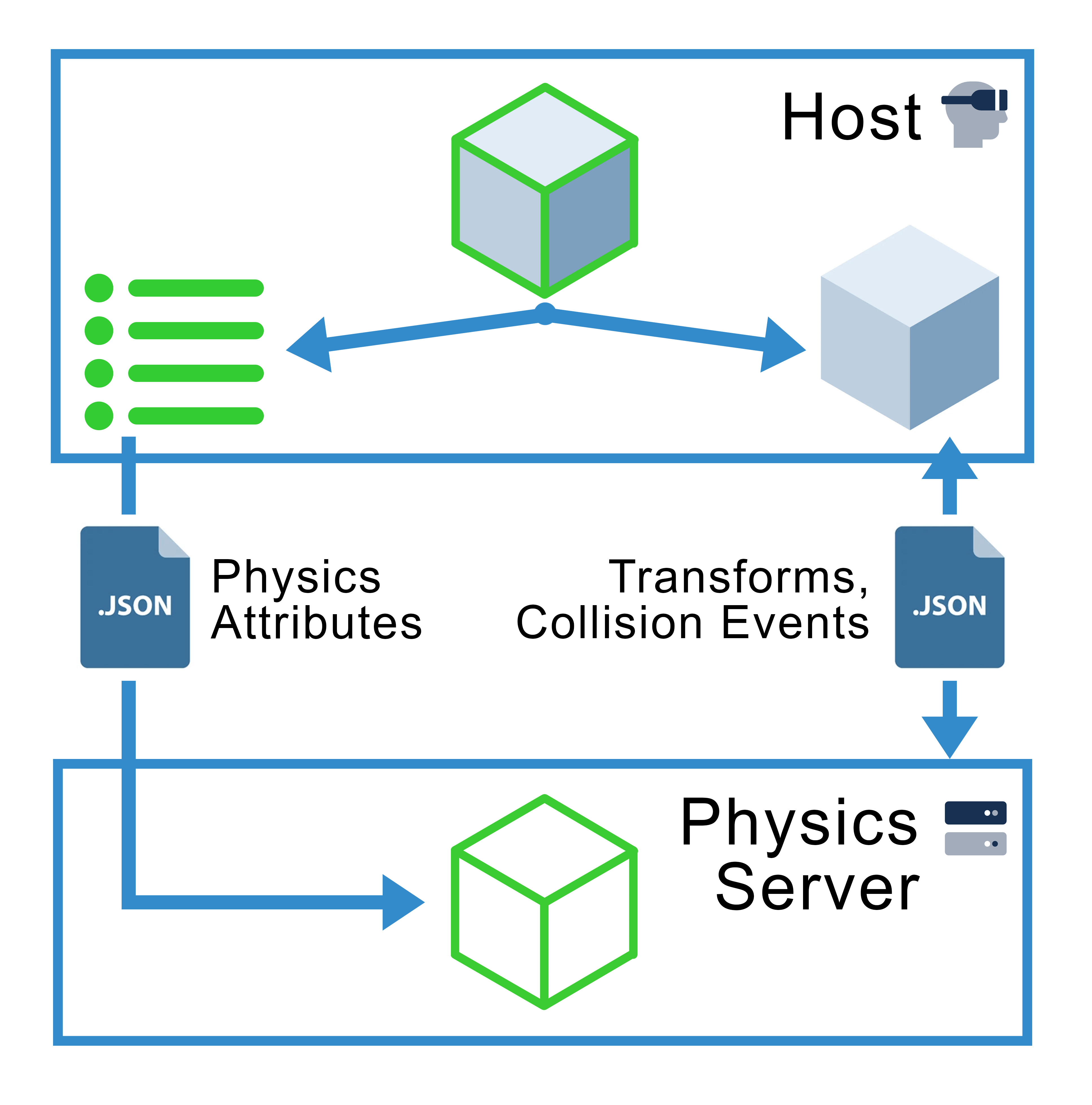}}
\caption{A high-level architecture diagram of the physics 
dissection. The physics are calculated only on the physics 
server while the render service receives the rendered image.}
\label{fig:physicsDissectionDiagram}
\end{figure}

\begin{figure}
\centerline{\includegraphics[width=18.5pc]{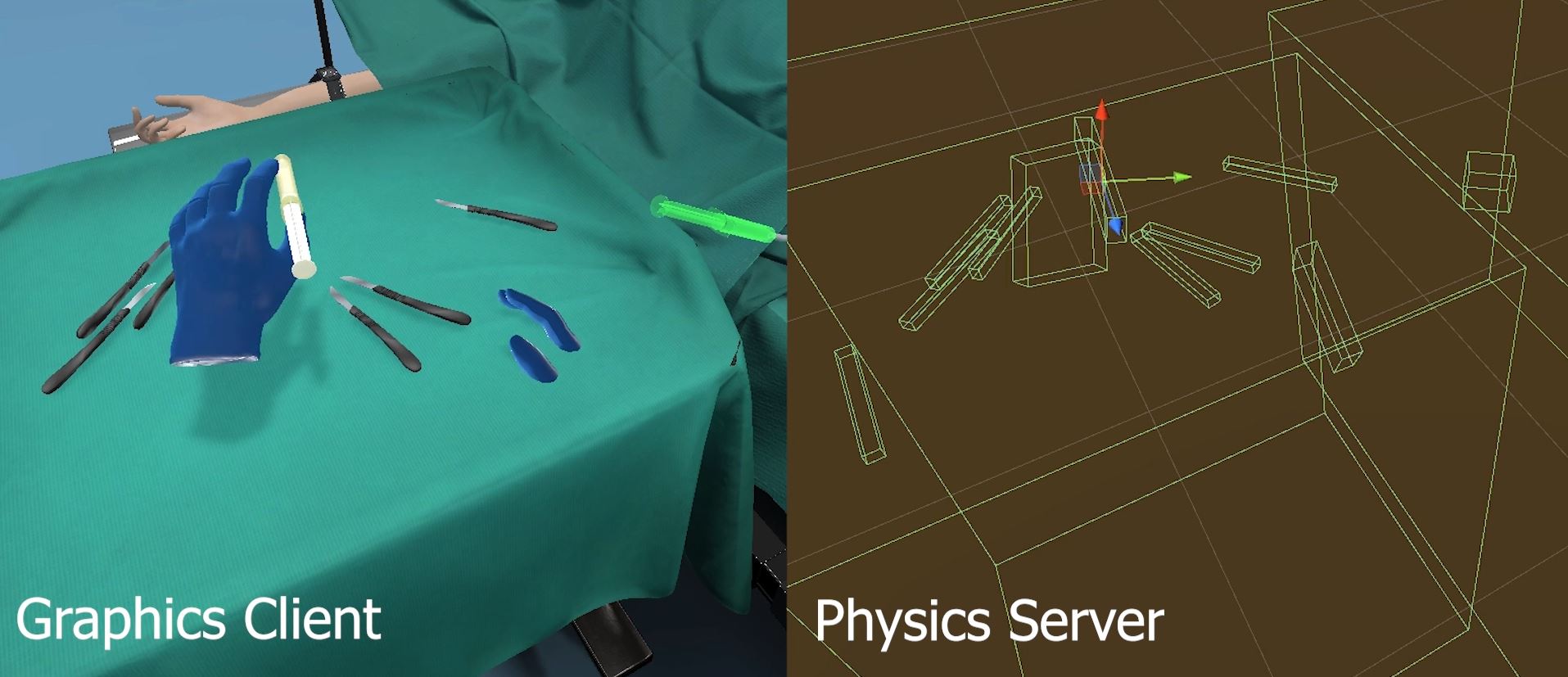}}
\caption{The physics are calculated only to the server while the client receives the rendered image.}
\label{fig:physicsRemoteDissection}
\end{figure}

The designed methods and techniques for the dissection of the 
Unity3D game engine pipeline, creates two autonomous, deployed 
separately, bidirectionally communicating components: the \emph{Host}, 
and the \emph{Physics Server}. The Host is responsible of 
maintaining the game logic and of processing the graphics 
rendering, while the Physics Server is responsible for 
performing physics computations
(see Fig.~\ref{fig:physicsDissectionDiagram}). 
The main goal of the dissected Unity3D pipeline is to allow 
any game object on the Host’s scene to be fully simulated by 
the separated Physics Server (see 
Fig.~\ref{fig:physicsRemoteDissection}). 

The dissection approach is based on splitting the 
Host's game objects into graphics objects, 
residing in the Host, and physics objects, residing in 
the Physics Server. This is accomplished by 
eliminating the physics attributes (e.g., mass, collider 
dimensions etc.) from all Host's game 
objects, which are subsequently created in the Physics 
server as physics objects with the same 
parameters, in an easily transmittable and compressible format. 
The Physics Server acts on a completely passive nature since 
it retains no knowledge regarding the Host’s scene, the game 
loop or the behaviors. The imposed inter-calls are streams of 
transformations between the two services.

The dissected Physics Server runs inside a Windows VM on a linux host. 
Further containerization of this service is under consideration. 
The results of the dissected pipeline, 
show a minor uptake of 0.03ms on the latency from the new physics 
service, also producing a frame-rate of \gr{60 fps}. 
The produced results have confirmed our plans of 
hosting extremely high-intensity physics computations in a 
separate edge service, and allow the Physics server to 
serve multiple Rendering services, collaborating 
in the same multi-user session.

\section{VR SOFT-BODY, SKINNED SIMULATION}

In the real world, there exist certain deformable objects, e.g., 
soft or hard tissues, which are ``naturally'' deformed 
when external forces are applied on them. 
MAGES incorporates a framework that aims to 
achieve this, via the so-called 
\emph{soft-body mesh deformation}. 
The idea behind this methodology is to create a 
layer of control points, called \emph{particles}, 
on top of the mesh model that, when translated, would affect 
the vertices of a model in a weighted manner. Ultimately, the 
visual effect we aim for is similar to what would happen in 
reality if we pinched and pulled the material at that point, 
towards the displacement direction of the particle. 

The particles are spawned and bound via springs to some initial 
positions on the mesh's surface, uniformly distributed 
via a Poisson Sampling mechanism. Particles 
lying sufficiently close are ``connected'', forming the desired 
layer of control points. Furthermore, we assign vertices to 
particles that have a distance from them below a specific 
model-dependent threshold.

After this initial setup, every time a particle
is displaced away from 
its initial position it will trigger several 
events. Firstly, the particle's displacement 
will amount to a weighted displacement of all particles 
it is assigned too, with each vertex's displacement 
being inversely proportional to its distance from the particle.
Furthermore, the particle's movement will affect  
connected particles by a fraction also inversely proportional
to their in-between distance. As neighbouring particles 
are displaced, they will, in turn, affect vertices in 
a diminishing extent. Lastly, particles that are moved away 
from their initial position tend to return with a velocity 
that is proportional to the displacement, similarly to 
a spring. With these fundamental rules applied to 
the model, we are able to simulate elasticity on the 
model's surface.

\begin{table}[tbp]
  \caption{Running times required to tear and cat various softbody models: a liver and a heart. Results taken from \cite{ProgressiveTearing}.
  }
  \begin{center}
  \begin{tabular}{|c|c|c|}
  \hline
  Characteristics & Liver & Heart \\
  \hline
  \hline 
  Number of vertices & 515 & 2527 \\
  Number of triangles & 768 & 4968 \\
  Number of particles & 191 & 179  \\
  \hline
  \hline 
  Tear Operation & \multicolumn{2}{c|}{Performance per tear segment} \\
  \hline
  \hline
  Perform Tear & 0.4 ms & 3 ms  \\ 
  Update particles & 2.3 ms  & 7.3 ms \\
  \hline
  Total Time & 3.25 ms & 11.19 ms  \\
  Output & 140 fps & 90 fps  \\
  \hline
  \hline 
  Cut Operation & \multicolumn{2}{c|}{Performance} \\
  \hline
  \hline
  Intersection points & 128 & 356  \\ 
  \hline
  Total Time & 10.9 ms & 17.2 ms  \\
  Output & 90 fps & 55 fps  \\
  \hline
  \end{tabular}
  \label{tbl:ctd_running_times}
  \end{center}
\end{table}


\section{HIGHLY REALISTIC PROGRESSIVE TEARING AND CUTTING}

By exploiting the GA-based interpolation engine, we are able to 
simulate realistic unconstrained consecutive tears or cuts, 
on a soft-body model, similar to the ones 
performed in real life by a surgeon in the operating room. 
Based on pure geometric operations 
on the surface mesh, we are able to perform such actions 
and obtain real-time results in XR, even in low-spec devices 
such as mobile VR HMDs.

Although \gr{Wu et al.} \cite{CUT2} describes diverse ways 
on how to cut a 3D model, most of these methods are not 
suitable for VR \cite{SurgicalCutting}, since the necessary computations must be performed in real-time, within a few ms 
to preserve user immersion. 
Latest developments \cite{ProgressiveTearing} allow for complex operations, 
such as cutting or tearing on a rigged mesh model, 
to be run in real-time, and are incorporated in MAGES. 
The significance of this framework lies on the fact that it 
overcomes current state-of-the-art limitations, 
where similar tears on a rigged 3D model in VR are 
predefined via linear-blend skinning animations, in order to allow 
them to playback in real-time.

Paired with the soft-bodies framework, our tearing and cutting 
algorithms allow the simulation of realistic, surgical-grade
continuous tears, especially valuable in the context of medical VR 
training. Furthermore, our algorithms are based on 
simple geometric predicates on the rigged 
mesh, and therefore do not require specific model pre-processing. 
Ultimately, our framework allows the user to freely cut
or tear in a consecutive way any 3D mesh model, 
under collaborative networked VE, reaching up to 140 fps,
depending on scene complexity (see Table~\ref{tbl:ctd_running_times}). 
Tears can now be simulated to 
``open'', replicating the tissue behaviour in real-life incisions, 
providing immersive visual results for soft-body materials.
And of course, rigged models can be further re-animated and 
torn or cut again, enabling a number of complex surgical 
operations to be implemented in XR.

\begin{figure}
\centerline{\includegraphics[width=18.5pc]{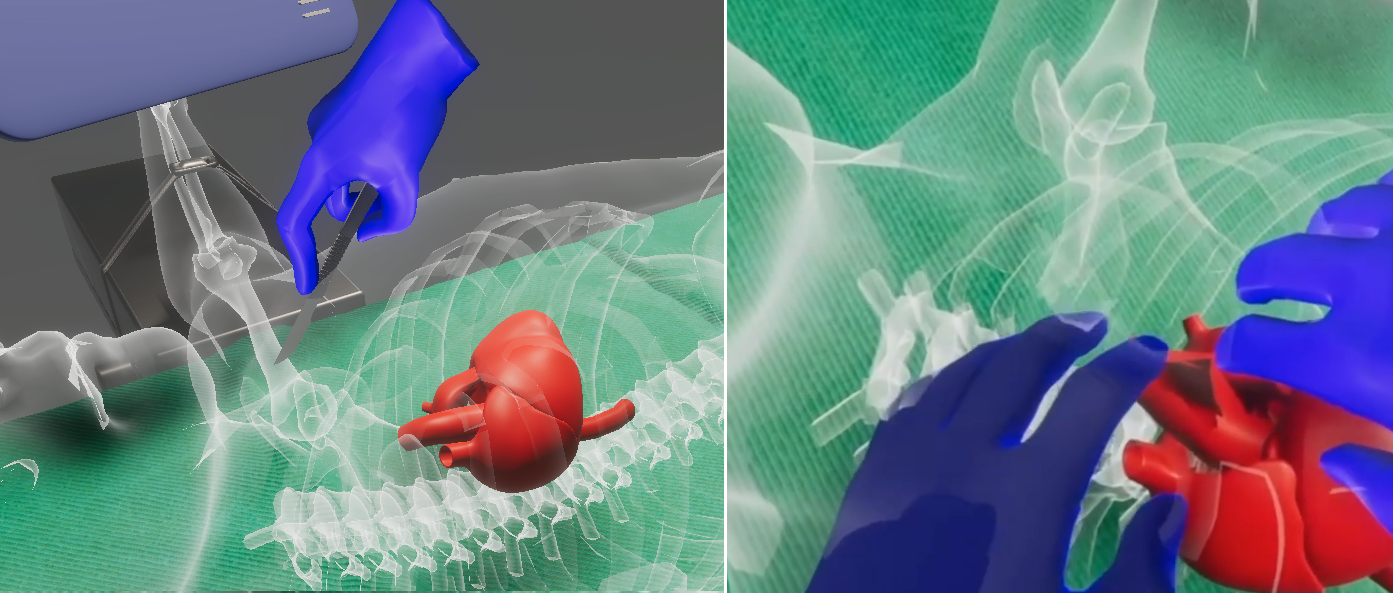}}
  \caption{Interactive heart model featuring our deformable and tearing algorithms. The student performed an incision to observe the internal structures.}
  \label{fig:tearingHeart}
\end{figure}

\section{CONVOLUTIONAL NEURAL NETWORK AUTOMATIC ASSESSMENT}
Despite the effort, only \gr{few} systems involve procedures 
for assessing user progress inside the immersive environment, 
that either evaluate only trivial tasks or require a huge 
amount of time by manually being reviewed or assessed. 
On the other hand, the need for real-time automated evaluation 
of user’s actions is constantly increasing. 
State-of-the-art methods for similar tasks either require 
the development of complicated task specific computer 
vision algorithms or support very simple tasks \cite{DL3}.

MAGES proposes a deep learning based system, that is able 
to assess, in real-time, user actions within a VR training scenario. 
The method enables the rapid development of trained assessment 
functions, since it utilizes data augmentation to minimize the 
amount of labelled data (e.g., poor/mediocre/excellent performance) 
that need to be collected. 
Using this system, we are able to assess actions 
(e.g., tears and cuts), performed in VR medical operations 
by the trainees. 

\begin{figure}[b]
    \centering
    \includegraphics[height=60mm, width=0.47\textwidth]{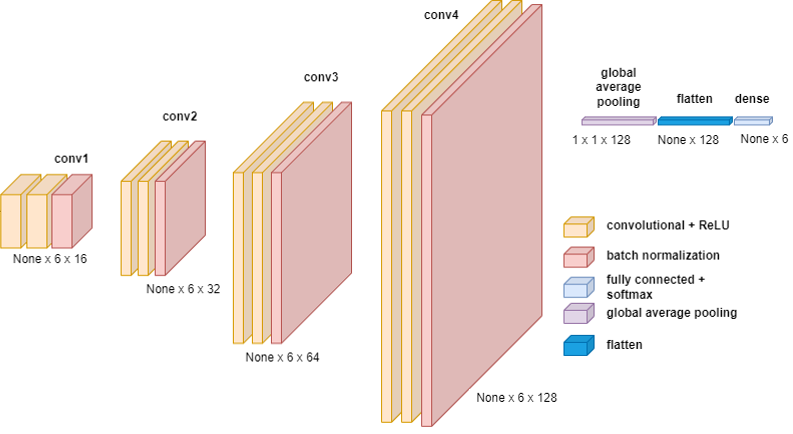}
    \caption{The layers of the proposed CNN model.}
    \label{fig:cnn}
\end{figure}

\gr{The developed scoring system evaluates user's actions, 
based on the trajectories of the used virtual tool. 
We utilize a supervised learning process that feeds score-labelled 
trajectories, randomly sampled, to a lightweight $15$-layer CNN model, 
able to provide high-accuracy results (see Fig.~\ref{fig:cnn}). 
The CNN consists of $4$ sets of two Convolutional, two ReLU, 
and a Batch Normalization layer, along with Global Average 
Pooling, Flattening and Densing layers, which form the classifier.
The model outputs the probabilities $p(i)$ of a trajectory 
belonging to each of the used classes. 
In Table~\ref{tbl:comparison}, a comparison of the accuracy 
of different models is presented for various number of classes. 
In the case of $2$ or $3$ classes, all models perform correctly, 
however, when $6$ classes are used, only our proposed CNN method 
achieves an acceptable performance. The model was trained with the 
VR Recorder's exported transformations/trajectories data, that were 
obtained after recording and labelling multiple sessions.}

\begin{table}[tb]
    \centering
    \caption{Table comparing the accuracy 
    of traditional ML techniques and the proposed CNN.}
    \begin{tabular}{|c||c|c|c|c|}
        \hline
        No. of & Logistic & \multirow{2}{*}{KNN} & \multirow{2}{*}{SVM}  & \multirow{2}{*}{CNN (ours)}   \\
        Classes & Regression & & & \\
        \hline\hline 
        2 & $100\%$ & $100\%$ & $100\%$ & $100\%$\\ \hline
        3 & $100\%$ & $100\%$ & $100\%$ & $100\%$\\ \hline
        6 & $39\%$ & $66\%$ & $83\%$ & $\textbf{95\%}$\\ \hline
    \end{tabular}
    \label{tbl:comparison}
\end{table}

The usefulness of this feature is especially highlighted 
in cases where we have a predefined 
operation, or exemplary task from a teacher, and students 
must be trained to replicate the teacher's movements precisely. 
Furthermore, by using transfer learning, these assessment 
functions can be reconfigured to support similar tasks, 
thus reducing even more the amount of training data. 

Various underlying techniques were used to make this achievable. 
Data collection involved capturing transformation data 
(translation \& rotation) of the active virtual tool 
(e.g., a scalpel) on distinct frames. The data set was 
made uniform to amend for the fact that the execution of an 
action is user dependent. Data augmentation techniques were
applied to increase the training data towards enhancing 
the training process of the $15$-layer CNN. 
Since low training and inference times are preferred, 
the lightweight model used is able to provide high-accuracy 
scoring results. Incorporating the trained CNN in the MAGES allows 
assessing the user's actions with minimal performance overhead.

\section{VR RECORDER FOR AUTOMATIC ASSESSMENT AND DEBRIEF}
Recording and replaying a training session in VR can serve as an 
additional and powerful educational tool. 
The bibliography shows great interest on this matter 
with the majority of projects focusing on video recording and motion capturing \cite{DL3}.

MAGES incorporates a different, highly accurate and lightweight solution to record and replay any collaborative training VR/AR 
session \cite{REC}. 
The underlying algorithm allows for the first time to record the 
session by capturing the transformation of objects 
(position, rotation) and user-driven events (interaction, decisions). 
This results in a highly accurate recording of the scenario that 
requires minimal storage space (approximately 1MB per minute per user) 
and minimal performance overhead (see Table~\ref{tbl:recorder_performance}). 
Additionally, we efficiently handle and record the user audio 
and synchronize it using a timestamp algorithm. 

After recording the session, users can select the recordings and replay them. They can join in the same recorded session,
not only as viewers of a three dimensional video but with the 
ability to navigate within the scene and re-live the session from any perspective in space, at any given time. In this way, they may
pay attention to details that 
they missed when they initially ``played'' the scenario.

\begin{table}
  \caption{Measuring the FPS burden of a VR application due to 
  the Recorder feature. Results taken from \cite{REC}. }
  \begin{center}
  \begin{tabular}{|c|c|c|c|}
  \hline
  \multirow{2}{*}{Performance}  & Session without & Session with   \\
  & VR Recording & VR Recording   \\
  \hline
  \hline
  Average FPS & 89.56  & 85.13  \\
  \hline
  Minimum FPS& 76.56  & 68.78  \\
  \hline
  Maximum FPS& 93.29  & 92.57  \\
  \hline 
  \end{tabular}
  \label{tbl:recorder_performance}
  \end{center}
\end{table}

\section{MEDICAL METAVERSE CASE STUDIES}

\gr{We can create a complete simulation at a fraction of the time 
and cost with respect to existing solutions. To be more precise, we
were able to build a training simulation for soft skills in 
collaboration with the Fayetteville State University within 14 days, employing 1 full-time developer and 1 part-time designer. 
Using other frameworks, the respective work would require 120 
days with the same human resources, to be completed. 
This corresponds to an 8 times faster and 8 times more cost-efficient authoring 
process, based on current freelance rates for 1 3D designer and 1 3D developer in USA today. For more simulation applications built with MAGES, 
along with the required amount of resources, the reader may refer 
to ``\nameref{sidebar:case_studies}''. }

\section{CONCLUSIONS AND FUTURE WORK}
In this work we presented MAGES 4.0, a novel development kit 
that allows rapid creation of any collaborative medical VR/AR 
simulation. The above novelties, rapidly accelerate content 
creation while maintaining a stable, multi-plaftorm authoring 
tool for the upcoming medical metaverse.

In the near future, we aim to improve the developer and user experience in 
MAGES. Our goal is to reach a point where a novice developer 
can rapidly uptake and create with all existing actions, cloud and 
analytics in a single day. In addition, experienced developers 
should also be able to extend the code-base (Actions, prototypes, mechanics) 
with ease at a fraction of time with minimal effort. Furthermore, we consider to support the Universal 
Scene Description (USD), an open metaverse file-format that empowers the 
collaboration in 3D virtual worlds. Our hope is that this work provides a solid contribution on how to accelerate world's transition to medical VR training and the proliferation of the medical metaverse.

\section{ACKNOWLEDGEMENT}
The project was partially funded by the European Union’s Horizon 
2020 research and innovation programme under grant agreements 
No 871793 (ACCORDION), No 101016509 (CHARITY) and No 101016521 
(5G-EPICENTRE) and from the Innosuisse - Swiss Innovation Agency   No 100.133 IP-ICT (Intelligent Digital Surgeon).

\newpage
\setcounter{figure}{0}
\renewcommand{\thefigure}{S\arabic{figure} }

\section[Medical VR Training Examples Built with MAGES]{Sidebar: Medical VR Training Examples Built with MAGES}
\label{sidebar:case_studies}

In this section we present honorable medical simulations built with MAGES SDK.

\begin{figure}[b]
\centerline{\includegraphics[width=18.5pc]{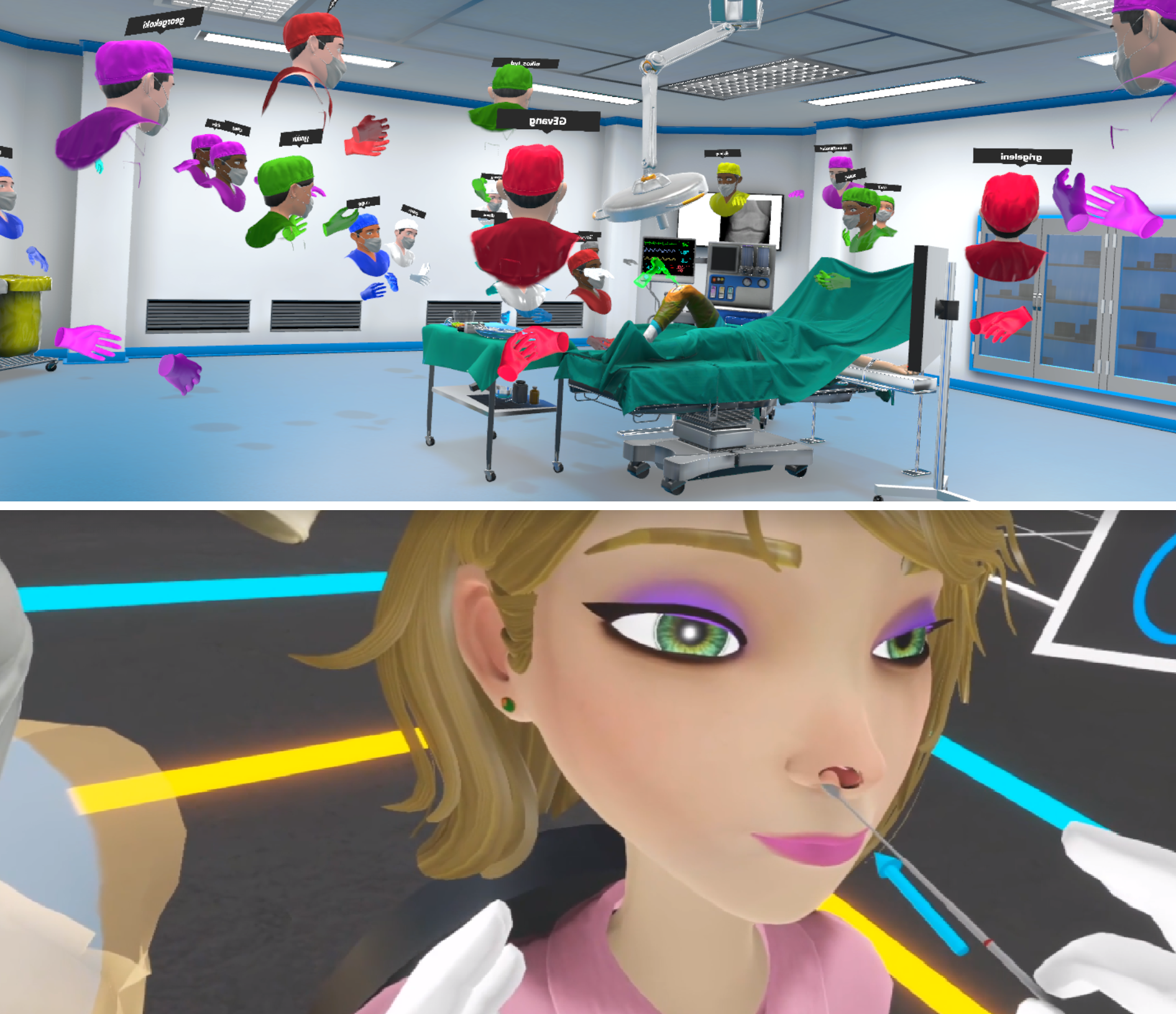}}
\caption{(Top)A MAGES 4.0 collaborative total knee arthroplasty simulation with 25 trainees in the same VR operating room. (Bottom) The MAGES 4.0 swab insertion process from the covid-19 nasopharyngeal swab test and Personal Protective Equipment simulation.}
\label{fig:casestudies1}
\end{figure}

\begin{figure}[b]
\centerline{\includegraphics[width=18.5pc]{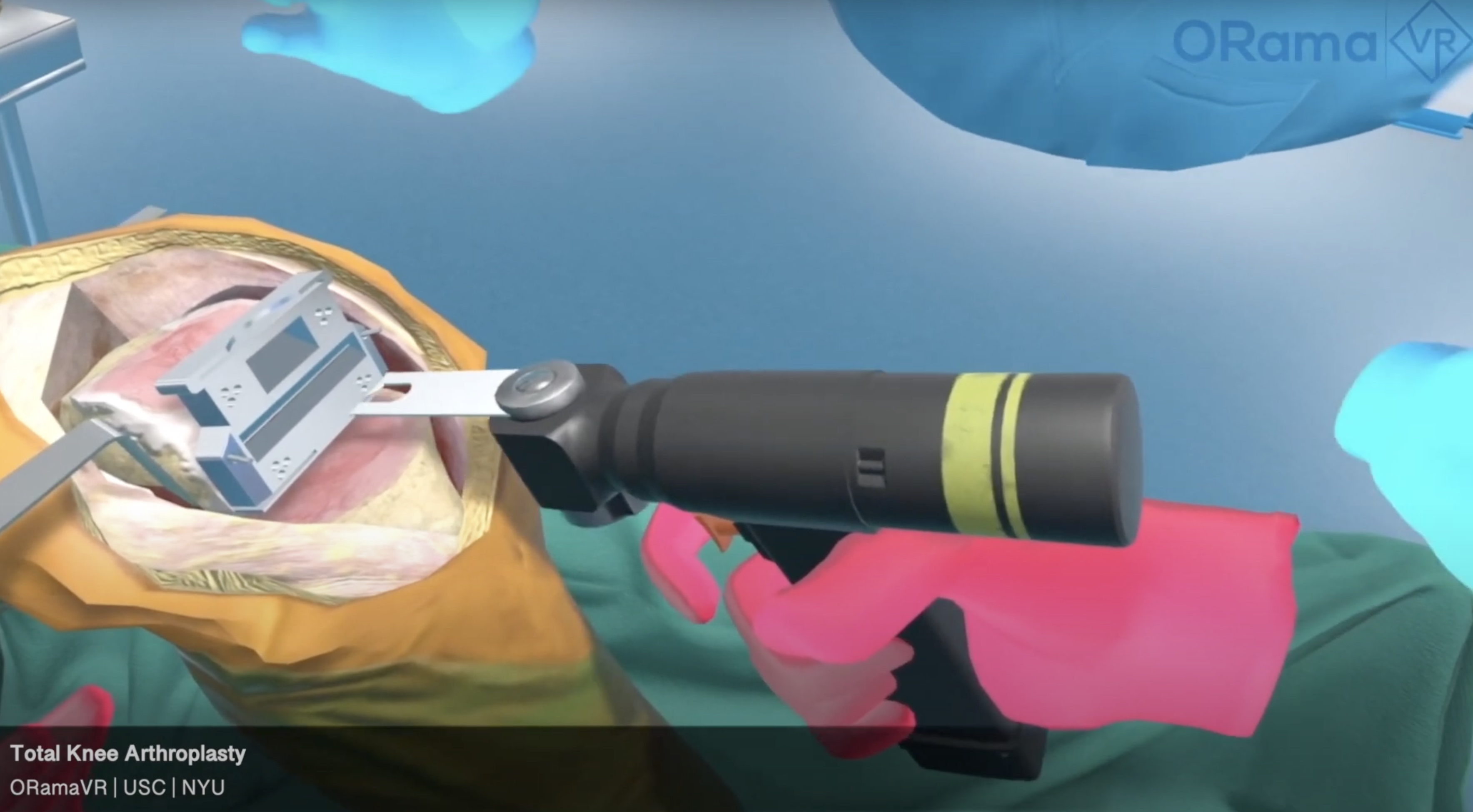}}
\caption{A MAGES 4.0 Total knee arthroplasty simulation.}
\label{fig:casestudies2}
\end{figure}

\begin{figure}[b]
\centerline{\includegraphics[width=18.5pc]{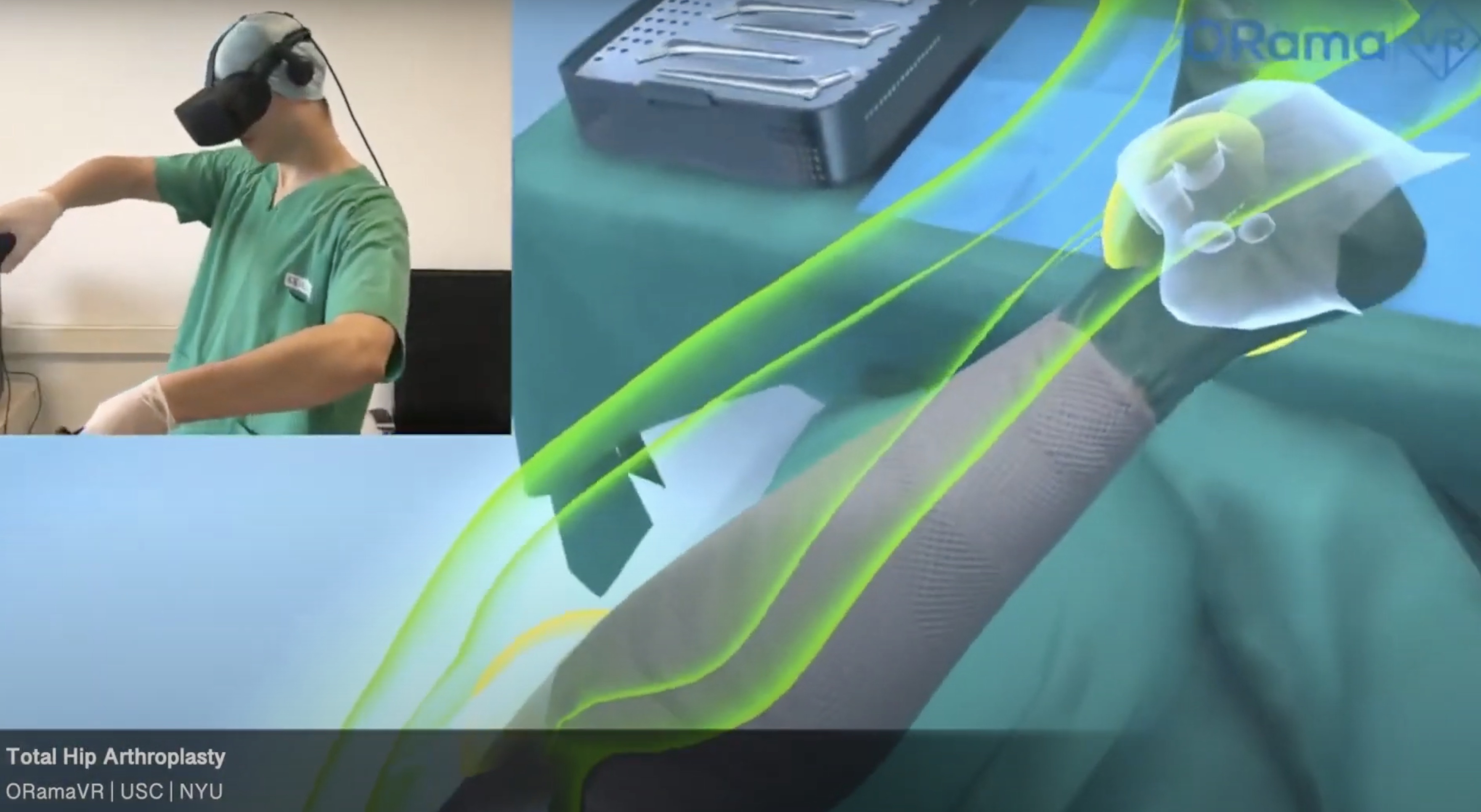}}
\caption{A MAGES 4.0 Total hip arthroplasty simulation.}
\label{fig:casestudies3}
\end{figure}

\textbf{Total knee and hip arthroplasty:} In collaboration with University of Southern California and New York University, we created both orthopaedic operations as a part of their curricula (see Figs.~\ref{fig:casestudies1},~\ref{fig:casestudies2},~\ref{fig:casestudies3}). We also conducted a clinical trial that proves skill transfer from virtual to the real world \cite{HooperTrial} \gr{(3 months, 3 full-time developer, 1 full-time designer)}.

\begin{figure}[b]
\centerline{\includegraphics[width=18.5pc]{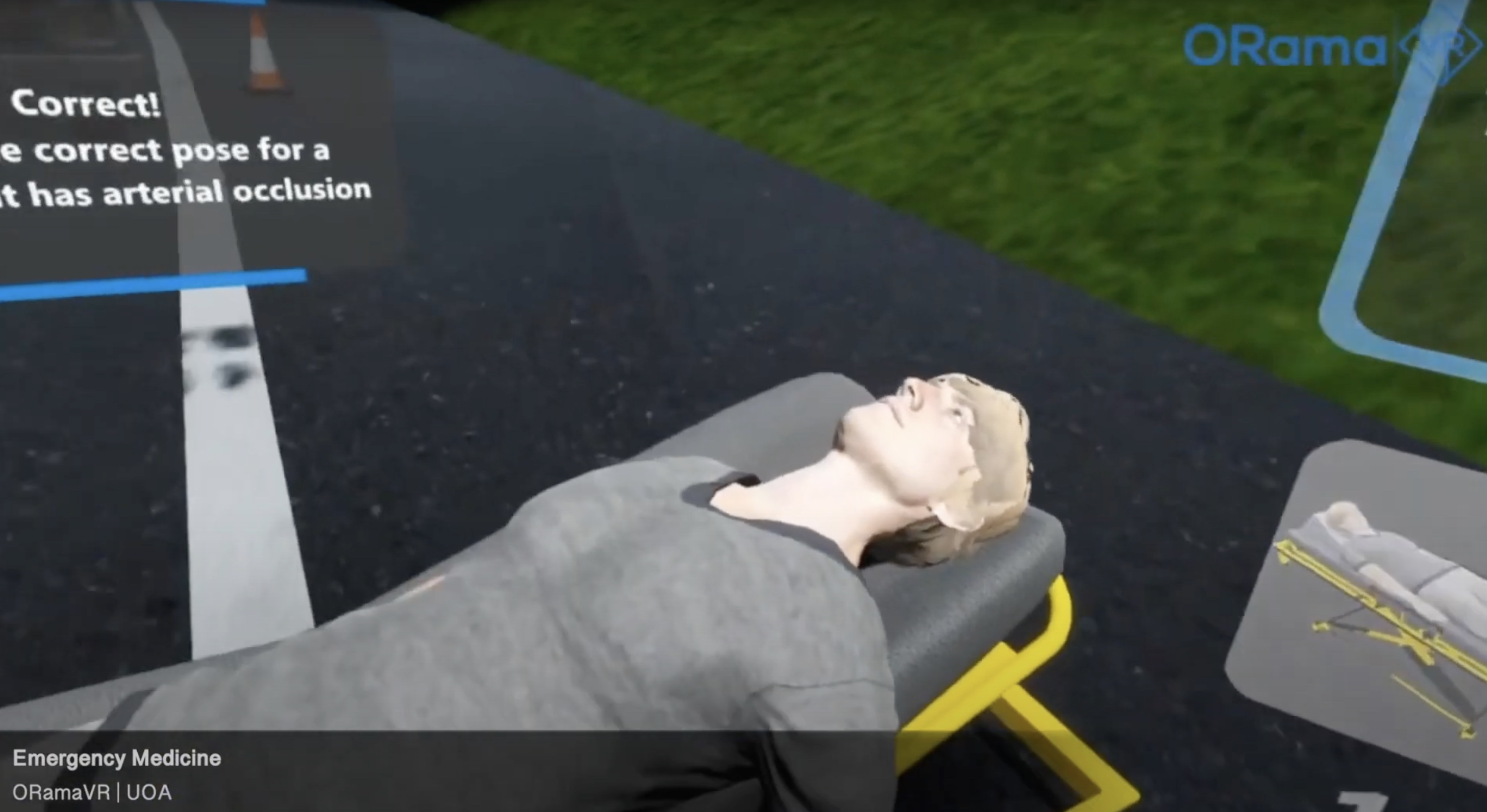}}
\caption{A MAGES 4.0 emergency trauma simulation.}
\label{fig:casestudies4}
\end{figure}

\begin{figure}[b]
\centerline{\includegraphics[width=18.5pc]{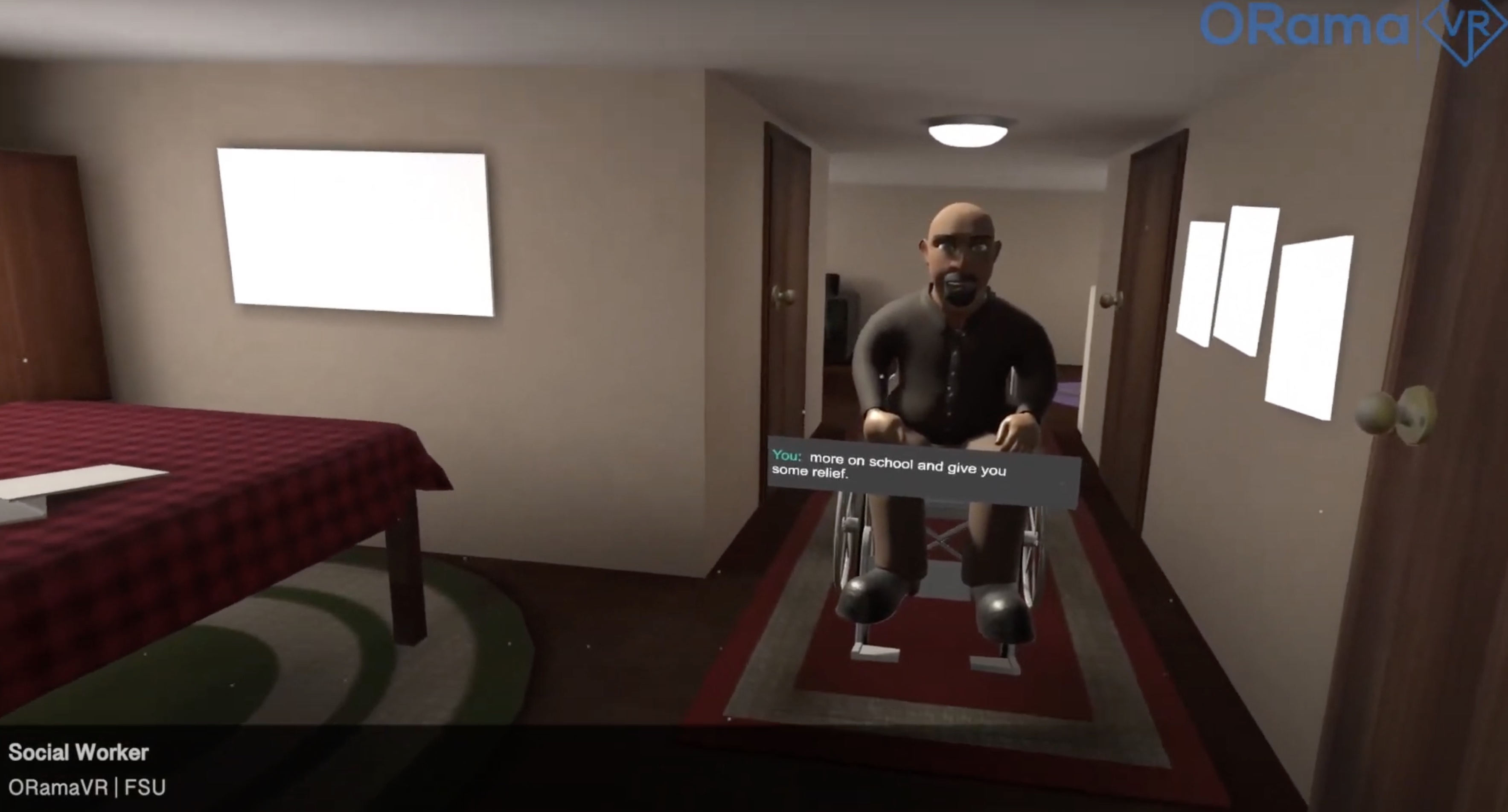}}
\caption{A MAGES 4.0 Social Worker simulation.}
\label{fig:casestudies5}
\end{figure}

\begin{figure}[b]
\centerline{\includegraphics[width=18.5pc]{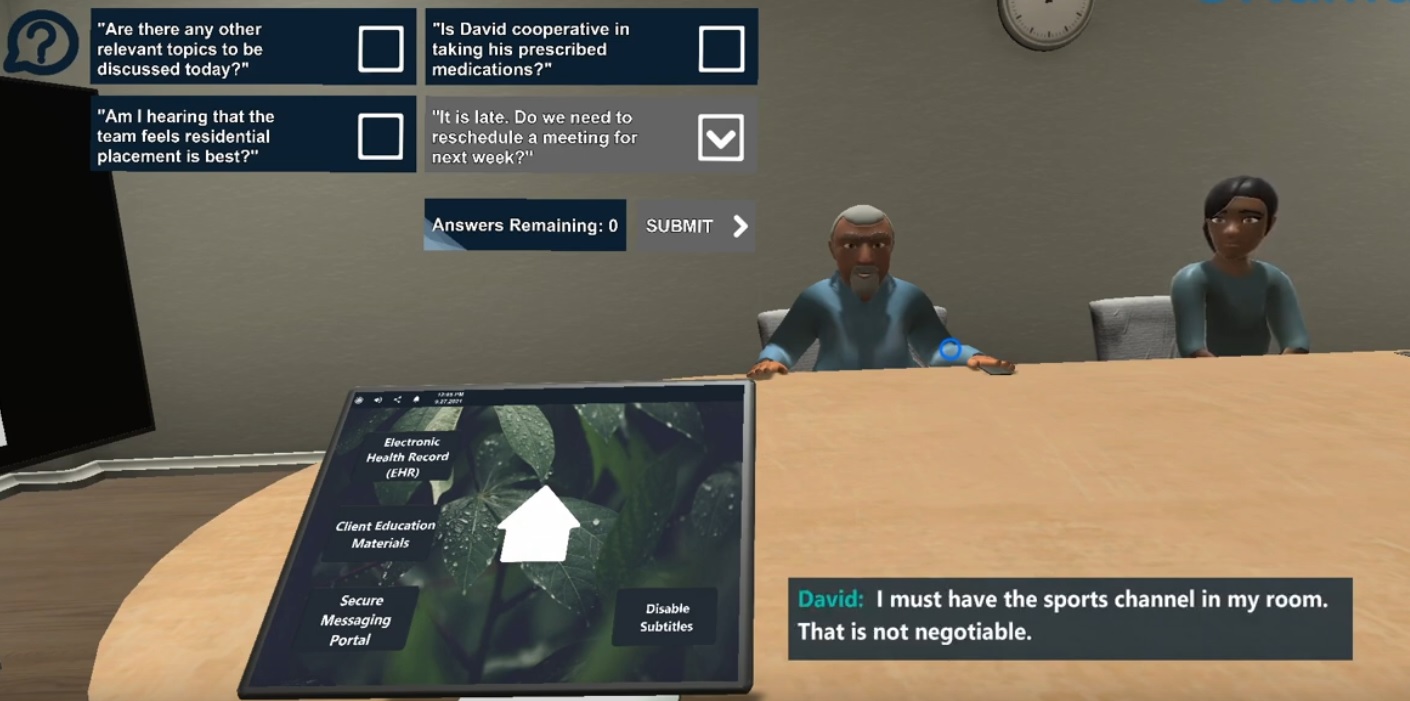}}
\caption{A MAGES 4.0 Behavioral Health simulation.}
\label{fig:casestudies5b}
\end{figure}

\begin{figure}[b]
\centerline{\includegraphics[width=18.5pc]{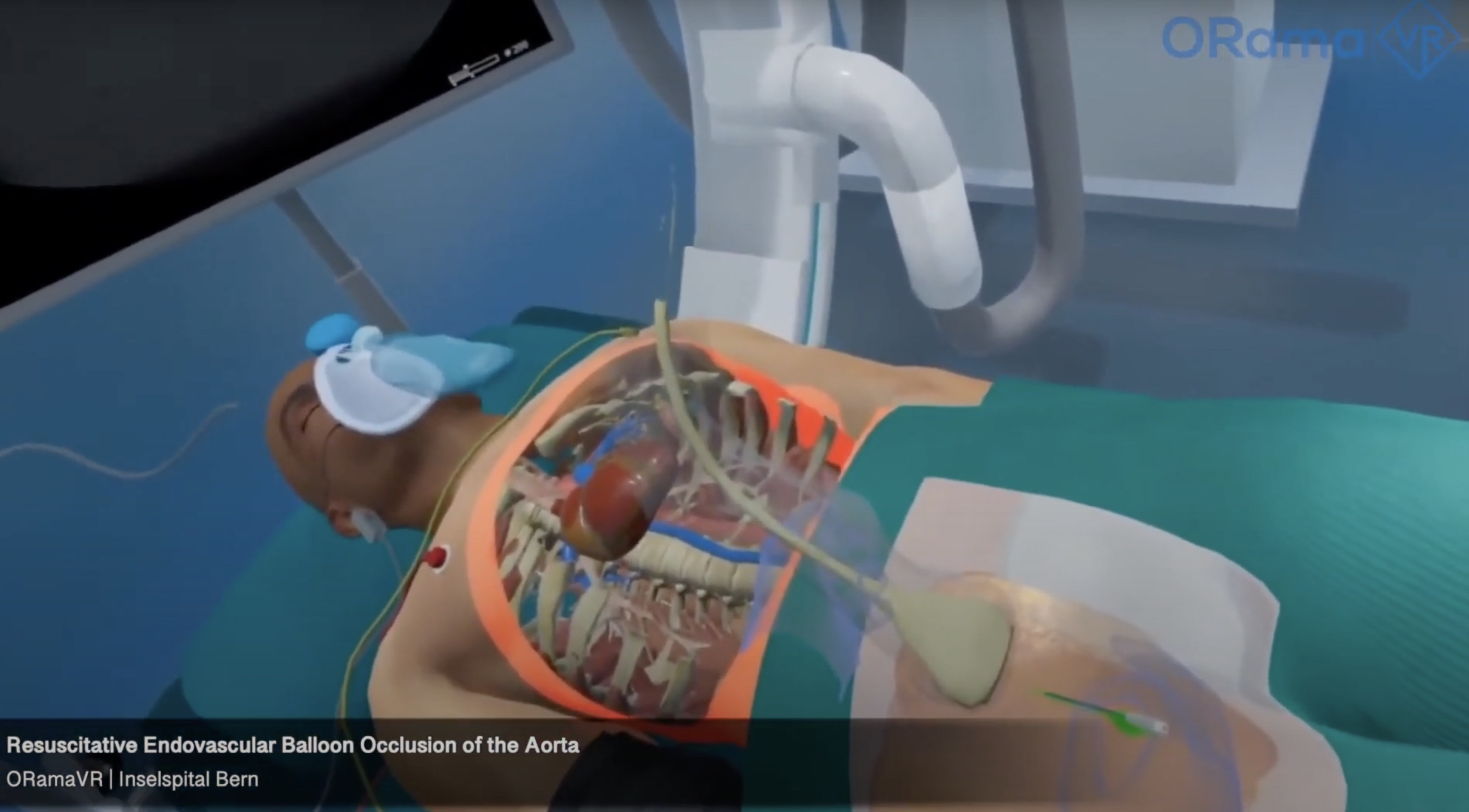}}
\caption{A MAGES 4.0 REBOA simulation.}
\label{fig:casestudies6}
\end{figure}

\begin{figure}[b]
\centerline{\includegraphics[width=18.5pc]{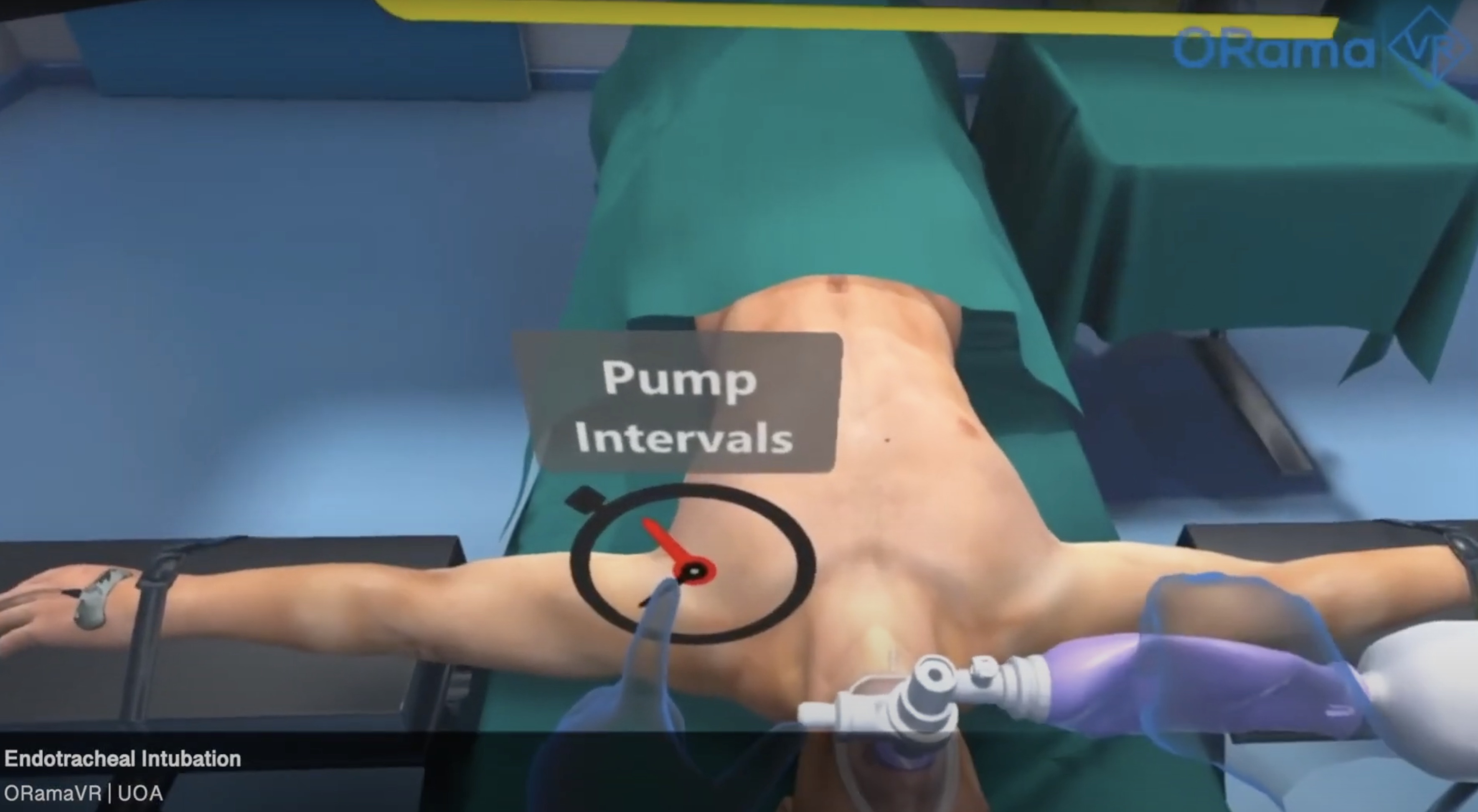}}
\caption{A MAGES 4.0 intubation simulation.}
\label{fig:casestudies7}
\end{figure}

\begin{figure}[b]
\centerline{\includegraphics[width=18.5pc]{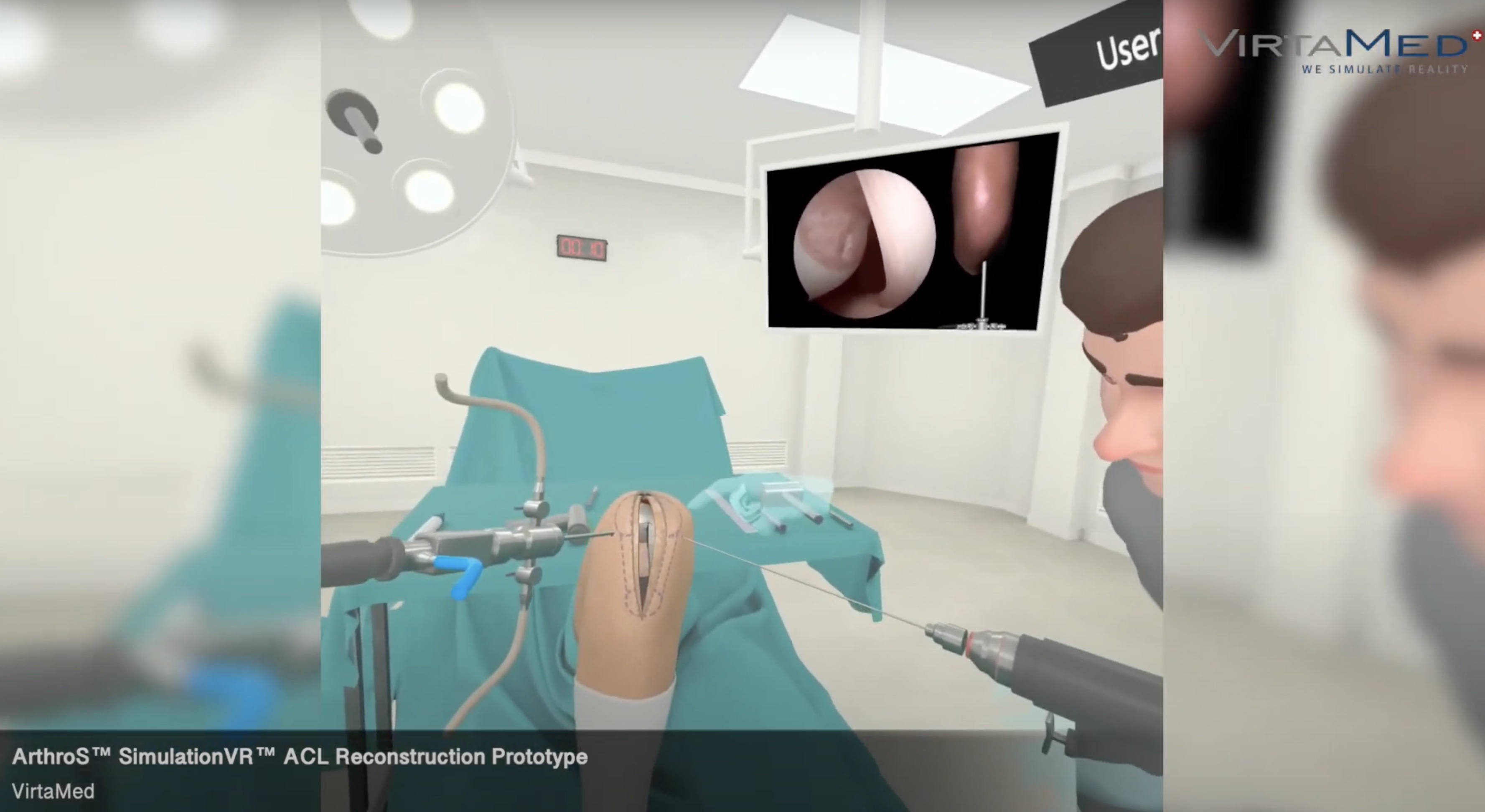}}
\caption{A MAGES 4.0 Arthroscopy simulation.}
\label{fig:casestudies8}
\end{figure}

\begin{figure}[b]
\centerline{\includegraphics[width=18.5pc]{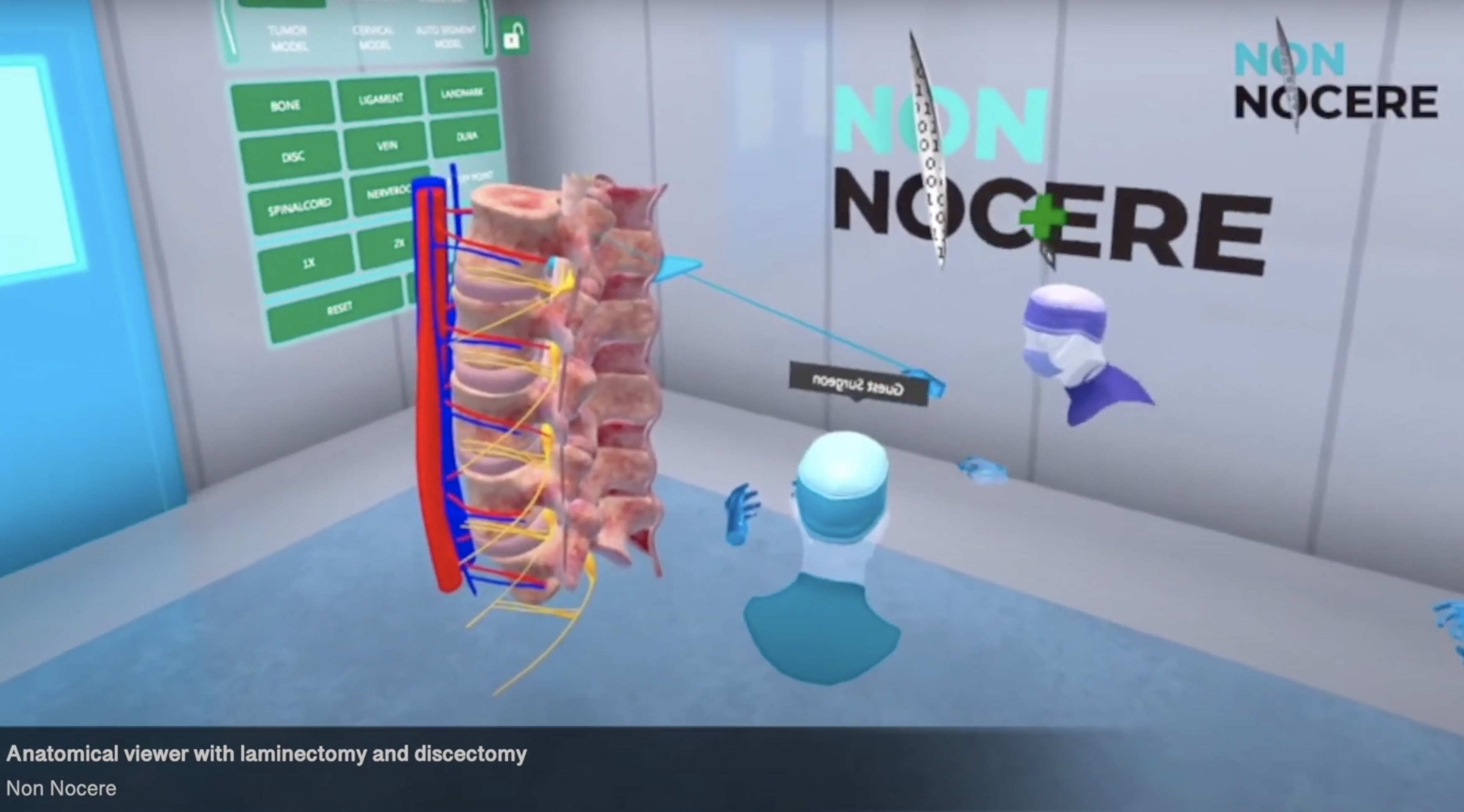}}
\caption{A MAGES 4.0 anatomical viewer simulation.}
\label{fig:casestudies9}
\end{figure}

\textbf{Emergency trauma scenarios:} This is a collection of emergency on-the-field simulations (extrication from car, first aid) and an intubation process (see Figs.~\ref{fig:casestudies4},~\ref{fig:casestudies7}) as part of an online course for the University of Athens \gr{(1 month, 2 full-time developer, 1 full-time designer)}.

\textbf{STARS:} Patient education and empowerment through knowledge. We developed this informative and stress relief application for patients in collaboration with ICS-FORTH and the Human Computer Interaction Lab \cite{STARS} \gr{(1.5 months, 2 full-time developer, 1 full-time designer)}.

\textbf{VRADA:} In collaboration with the Aristotle University of Thessaloniki, the University of Thessaly and Biomechanic solutions we created a VR bike simulation with cognitive questions that allows older people with mild cognitive impairment symptoms to simultaneously practice physical and cognitive skills on a dual task. We published a clinical trial with our results \cite{VRADA1} \gr{(2 months, 1 full-time developer, 1 full-time designer)}.

\textbf{Social Worker:} A cognitive training simulation for soft skills in collaboration (see Fig.~\ref{fig:casestudies5}) with the Fayetteville State University respectively \gr{(14 days, 1 full-time developer, 1 part-time designer)}.

\textbf{Behavioral Health, Chronic Care assessment:} Two cognitive training simulations for soft skills in collaboration (see Fig.~\ref{fig:casestudies5b}) with the Western Governors University \gr{(3 months, 2 full-time developers, 1 part-time designer)}.

\textbf{Covid-19 PPE \& swab testing:} During the covid-19 pandemic, we developed this application to educate medical personnel the proper use of personal protective equipment and how to take covid samples for test (see Fig.~\ref{fig:casestudies1}). It was performed in collaboration with Inselspital University hospital of Bern and the New York University. We conducted two clinical trials \cite{CovidTrial, CovidTrial2}, to explore the effectiveness of VR simulations versus traditional learning methods \gr{(3 months, 3 full-time developer, 2 full-time designer)}.

\textbf{Anatomical viewer:} Non Nocere, developed an anatomical viewer for laminectomy and discectomy using MAGES SDK (see Fig.~\ref{fig:casestudies9}).

\textbf{REBOA:} In collaboration with the Inselspital University hospital of Bern we developed a endovascular balloon occlusion of the Aorta simulation  (see Fig.~\ref{fig:casestudies6}) \gr{(3 months, 3 full-time developer, 1 full-time designer)}.

\gr{
\textbf{Colaboration with VirtaMed AG.:} VirtaMed AG, is currently developing a Knee Arthroscopy simulation (see Fig.~\ref{fig:casestudies8}) using MAGES SDK\footnote{https://business.vive.com/uk/stories/solving-medical-training-challenges-present-and-future/}.}

\gr{
\textbf{IDS: }In collaboration with University of Geneva, University Hospital of Geneva (HUG), and MiraLAB Sarl, in the Swiss funded project Intelligent Digital Surgeon\footnote{https://innosuisse.miralab.com/} (IDS), we are currently advancing our ML assessment feature. 
The main objective of the project is to identify and analyze the immersed trainee’s behavioral model and provide personalized real-time feedback, assessment, and recommendations like a real surgical instructor.
A deep learning model will be derived to identify the trainee’s behavioral model, by recognizing and analyzing the trainee’s hand/arm gestures, and to assist the feedback decision engine, by providing personalized assessment, real-time feedback, instructions, and recommendations.}

\gr{
\textbf{PROFICIENCY: }We also participate in the multi-partner (the lead of the three clinical partners Kantonsspital St.Gallen, Centre Hospitalier Universitaire Vaudois and Balgrist University Hospital and VirtaMed AG, Microsoft Mixed Reality, AI Zurich Lab, and Atracsys LLC) Swiss funded, PROFICIENCY\footnote{https://surgicalproficiency.ch/} project, that aims to define a fully novel, standardized and proficiency-based surgical training curricula installed and demonstrated on two example surgical modalities, laparoscopy and arthroscopy, while fully generalizable to other interventions. }

\begin{IEEEbiography}{Paul Zikas,} is Vice President of engineering at ORamaVR. He earned his BSc, and MSc in Computer Science from the University of Crete. Currently he is developing his PhD research topics at the University of Geneva. Contact him at paul.zikas@oramavr.com 
\end{IEEEbiography}

\begin{IEEEbiography}{Antonis Protopsaltis,} is a lead research scientist at ORamaVR. He holds a Ph.D. in Computer Science. He is also a Special Teaching Fellow with the University of Western Macedonia. He specializes in Computer Graphics, and CAD methods. Contact him at antonis.protopsaltis@oramavr.com 
\end{IEEEbiography}

\begin{IEEEbiography}{Nick Lydatakis,} is head of platform  at ORamaVR.He graduated from Computer Science department, University of Crete where he obtained both this BSc and MSc on Computer Graphics. Contact him at nick.lydatakis@oramavr.com 
\end{IEEEbiography}

\begin{IEEEbiography}{Mike Kentros,} is lead SDK developer \& ML at ORamaVR. He earned his BSc, and MSc in Computer Science from the University of Crete. Currently, he is PhD candidate in the same university. Contact him at mike.kentros@oramavr.com 
\end{IEEEbiography}

\begin{IEEEbiography}{Stratos Geronikolakis,} 
 is lead SDK developer, responsible of multiplatform support and products distribution, in ORamaVR. He acquired his BSc and MSc in Computer Science of the University of Crete, specializing in computer graphics. Contact him at stratos.geronikolakis@oramavr.com
\end{IEEEbiography}

\begin{IEEEbiography}{Giannis Evangelou,} 
is a Lead VR Developer \& Level Designer at ORamaVR. He obtained his BSc in Computer Science, University of Crete, specializing in game design and computer graphics. Currently he is pursuing his MSc in Information Systems and Human-Computer Interaction. Contact him at giannis.evangelou@oramavr.com
\end{IEEEbiography}

\begin{IEEEbiography}{Achilleas Filippidis,}
is a Content and SDK Developer in ORamaVR. He holds a BSc in Computer Science from University of Crete, specialized in Computer Graphics, and continues his academic path with an MSc. Contact him at achilles.filippidis@oramavr.com
\end{IEEEbiography}

\begin{IEEEbiography}{Manos Kamarianakis,} is an SDK R\&D developer \& ML at ORamaVR, holding a Ph.D. in Applied Mathematics. He specializes in Computational Geometry, Geometric Algebra and Computer Graphics applications. Contact him at manos.kamarianakis@oramavr.com
\end{IEEEbiography}

\begin{IEEEbiography}{Dimitris Angelis,} is a Machine Learning developer at ORamaVR. He earned his BSc in Computer Science from the University of Crete specialized in classification algorithms. Contact him at dimitris.angelis@oramavr.com 
\end{IEEEbiography}

\begin{IEEEbiography}{Michail Tamiolakis,} is content and soft body developer at ORamaVR. He earned his BSc in Computer Science from the University of Crete specialized soft body simulations.  Contact him at michalis.tamiolakis@oramavr.com 
\end{IEEEbiography}

\begin{IEEEbiography}{Michael Dodis,} is a networking and Unreal engine developer at ORamaVR. He first came into contact with programming through Unity, then went on doing lower-level projects with computer graphics. Contact him at michael.dodis@oramavr.com 
\end{IEEEbiography}

\begin{IEEEbiography}{George Kokiadis,} is a MR developer at ORamaVR. He completed his BSc in 2021 with a Thesis focused on Mobile XR and continues his academic path with an MSc in Multimedia Technology. Contact him at george.kokiadis@oramavr.com 
\end{IEEEbiography}

\begin{IEEEbiography}{John Petropoulos,} 
is VR content developer in ORamaVR. He holds a BSc in Computer Science and is pursuing his MSc in computer graphics. Contact him at john.petropoulos@oramavr.com
\end{IEEEbiography}

\begin{IEEEbiography}{Steve Kateros,} 
is Vice President of Product Management in ORamaVR. He has acquired his BSc and MSc in Computer Science from University of Crete, specialising in Computer Graphics and Virtual Reality. He has worked as researcher in ICS-FORTH. Contact him at steve.kateros@oramavr.com
\end{IEEEbiography}

\begin{IEEEbiography}{Eleni Grigoriou,} 
Eleni Grigoriou, is Product Manager \& UI Design at ORamaVR. She earned her BSc in Computer Science from the University of Crete spesialized in Big-Data interaction in VR. Currently she is pursuing her MSc on computer graphics. Contact her at eleni.grigoriou@oramavr.com
\end{IEEEbiography}

\begin{IEEEbiography}{Maria Pateraki,} is R\&D executive at ORamaVR. She is also an Assistant Professor in Photogrammetry at the National Technical University of Athens and an affiliated Researcher of the CVRL at ICS-FORTH. Contact her at maria.pateraki@oramavr.com 
\end{IEEEbiography}

\begin{IEEEbiography}{George Papagiannakis,} is co-founder and CEO/CTO of ORamaVR. His academic credentials include serving as Professor of Computer Graphics at the Computer Science department of the University of Crete, Greece, as Affiliated  Research Fellow at the HCI lab at ICS-FORTH, where he leads the CG Group and as visiting Prof of CS at the University of Geneva. Contact him at george.papagiannakis@oramavr.com 
\end{IEEEbiography}

\end{document}